\pdfoutput=1
\documentclass[10pt,journal]{IEEEtran}
\ifCLASSOPTIONcompsoc
  \usepackage[nocompress]{cite}
\else
  \usepackage{cite}
\fi
\usepackage[table]{xcolor}
\usepackage{multirow}
\usepackage[T1]{fontenc}
\usepackage{multirow}
\usepackage{amsmath}
\usepackage{amssymb}
\usepackage{mdsymbol}
\usepackage{tabularx}
\usepackage{mfirstuc}
\usepackage{url}
\usepackage{xspace}
\newcommand{\Paragraph}[1]{\smallskip\noindent{\it \textbf{#1:}}}

\usepackage{graphicx,epstopdf}

\newcommand{\acronym}{\emph{NoiSense}\xspace}

\usepackage{adjustbox}

\usepackage{placeins}

\usepackage{array}
\newcolumntype{P}[1]{>{\centering\arraybackslash}p{#1}}
\usepackage{bm}

\usepackage{nameref}
\usepackage{mathtools}

\usepackage{multirow}
\usepackage{verbatim}
\usepackage{eucal}

\def\skipnoindent{\vskip0.1in\noindent}
\begin{document}

\title{Challenges and Opportunities in CPS Security: A Physics-based Perspective}
\author{Chuadhry Mujeeb Ahmed and Jianying Zhou \\ Singapore University of Technology and Design \\ mujeeb\_chuadhry,jianying\_zhou@sutd.edu.sg}

\IEEEtitleabstractindextext{%
\begin{abstract}

 The integration of cyber technologies (computing and communication) with the physical world gives rise to complex systems referred to as Cyber Physical Systems (CPS), for example, manufacturing, transportation, smart grid, and water treatment. Many of those systems are part of the critical infrastructure and need to perform safely, reliably, and securely in real-time. CPS security is challenging as compared to the conventional IT systems. An adversary can compromise the system in both the cyber and the physical domains. However, the unique set of technologies and processes being used in a CPS also bring up opportunities for defense. CPS security has been approached in several ways due to the complex interaction of physical and cyber components. In this work, a comprehensive study is taken to summarize the challenges and the proposed solutions for securing CPS from a Physics-based perspective. 
\end{abstract}

\begin{IEEEkeywords}
Cyber physical systems, CPS security, Physics-based attack detection.
\end{IEEEkeywords}}

\maketitle

\IEEEdisplaynontitleabstractindextext
\IEEEpeerreviewmaketitle
\def\heading#1{\par\medskip\textbf{\textit{#1:}}}

    
    
    


    
    
    


\section{Introduction} 
\label{Sec_Intro} %

Recent progress in technology is resulting in the digitization of the physical world and things around us. It is expected that communication and computing capabilities will soon be part of all the physical objects~\cite{rajkumar_insup_CPS_intro_2010}. The integration of cyber technologies (computing and communication) with the physical world gives rise to complex systems referred to as Cyber Physical Systems~(CPS). CPS has changed the methods that humans used to interact with the physical world. Some examples of CPS are manufacturing, transportation, smart grid, water treatment, medical devices and the Industrial Internet of Things~(IIoT)~\cite{CPS_thesis_ETH_2018_intro}. 
Many of those systems are part of the critical infrastructure, and need to perform safely, reliably, and securely in real time. This article discusses the security issues related to CPS.


A CPS consists of  Programmable Logic Controllers~(PLC), sensors, actuators, Supervisory Control, and Data Acquisition~(SCADA) workstation and Human Machine Interface~(HMI) that are interconnected via a communications network. The PLCs control a physical process based on the sensor measurements. The advances in communication technologies help to better monitor and operate CPS, but this connectivity also exposes physical processes to malicious entities on the cyber and physical domains. Recent incidents of sabotage on these systems~\cite{ukranian_case2016analysis,slay_miller_2008,stuxnet}, have raised concerns on the security of CPS~\cite{cardenas2009challenges}.

Challenges in CPS security are different as compared with the conventional IT systems, especially in terms of consequences in case of a security lapse. Attacks on CPS might result in damage to the physical property, as a result of an explosion~\cite{aurora_attack,German_steelmill_attack} or severely affect people who depend on critical infrastructure as was the case of recent power cutoff in Ukraine~\cite{ukranian_case2016analysis}. Data integrity is an important security requirement for CPS~\cite{Gollmann2016} and hence the integrity of sensor data should be ensured. Sensor data can either be spoofed in cyber (digital) domain~\cite{urbina_CCS2016limiting} or in physical (analog) domain~\cite{shoukry2015,drone_Son2015}. Sensors are a bridge between the physical and cyber domains in a CPS. Traditionally, an Intrusion Detection System (IDS) monitors a communication network or a computing host to detect attacks. However, physical tampering with sensors or sensor spoofing in the physical/analog domain may go undetected by the legacy IDS~\cite{shoukry2015}. 

In this article, we briefly introduce CPS using an example from the electric power and water treatment system, highlight the challenges and opportunities based on the physics of the systems. Detection techniques based on physics of the process against attacks on sensor reading have been proposed in recent studies~\cite{ahmed2019state_book,shoukry2015,yasser-2013,drone_Son2015,sensor_saturationAttack_infusionpump_usenix2016,sampling_race2016,walnut_acoustic_attack_mems_accelerometer}. An attacker who tries to defy rules of physics would also expose itself. An understanding of the physics of the process can help to secure a CPS~\cite{ahmed_QRS2017}. A mini-survey of the existing techniques is presented by highlighting the limitations of the previous works and proposed improvements. A device fingerprinting technique used for attack detection in CPS is explained before concluding the article.

\begin{figure*}[!htb]
    \centering
    \includegraphics[scale=0.17]{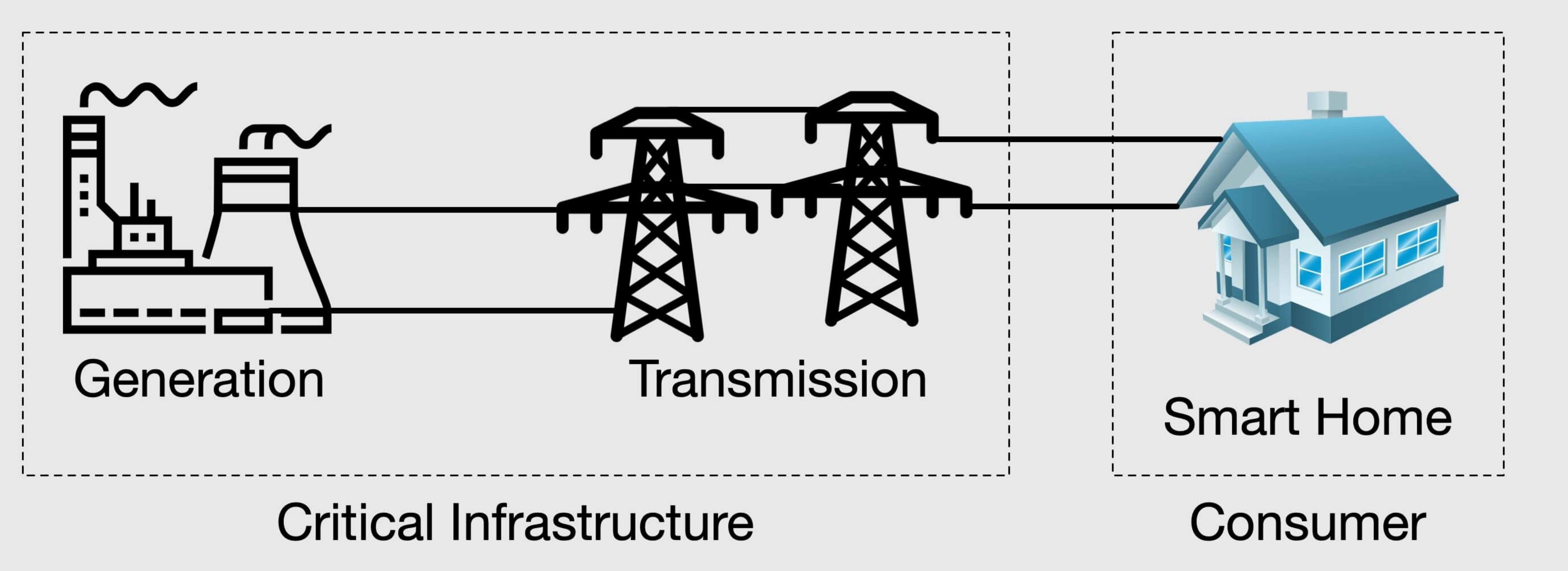}
    \caption{A generic electrical power system as an example of CPS.}
    \label{generic_electric_system_fig}
\end{figure*}

\section{Cyber Physical Systems}


Cyber Physical System (CPS) is a broad term for systems ranging from medical, power, transport and industrial systems. In the following we highlight two major sectors applicable to our daily life, that is, electrical power and water treatment systems. An example of a CPS is shown in Figure~\ref{generic_electric_system_fig}. It shows the high-level architecture of an electrical power system. This is composed of electricity generation (power plants), transmission (electric grid system) and end-users (smart home). As one can imagine this power system is composed of a multitude of devices and physical processes. Power generation and transmission depend on the demand from the utilities and the users. To meet the requirements of the energy demand the critical infrastructure is utilized to ensure a continuous supply of power.  Each of the processes in the critical infrastructure is a complex engineering system and needs a sophisticated control to achieve its desired objectives. For example, at the generation stage, we have generators, Intelligent Electronic Devices (IEDs) also incorporating electric relays, all these devices are autonomously controlled by the Programmable Logic Controllers (PLC). This means that we have a lot of sensors monitoring the physical process, actuators/generators and the physical infrastructure that communicate the current physical states with each other and with the PLC. 

\begin{figure*}[!htb]
    \centering
    \includegraphics[scale=0.17]{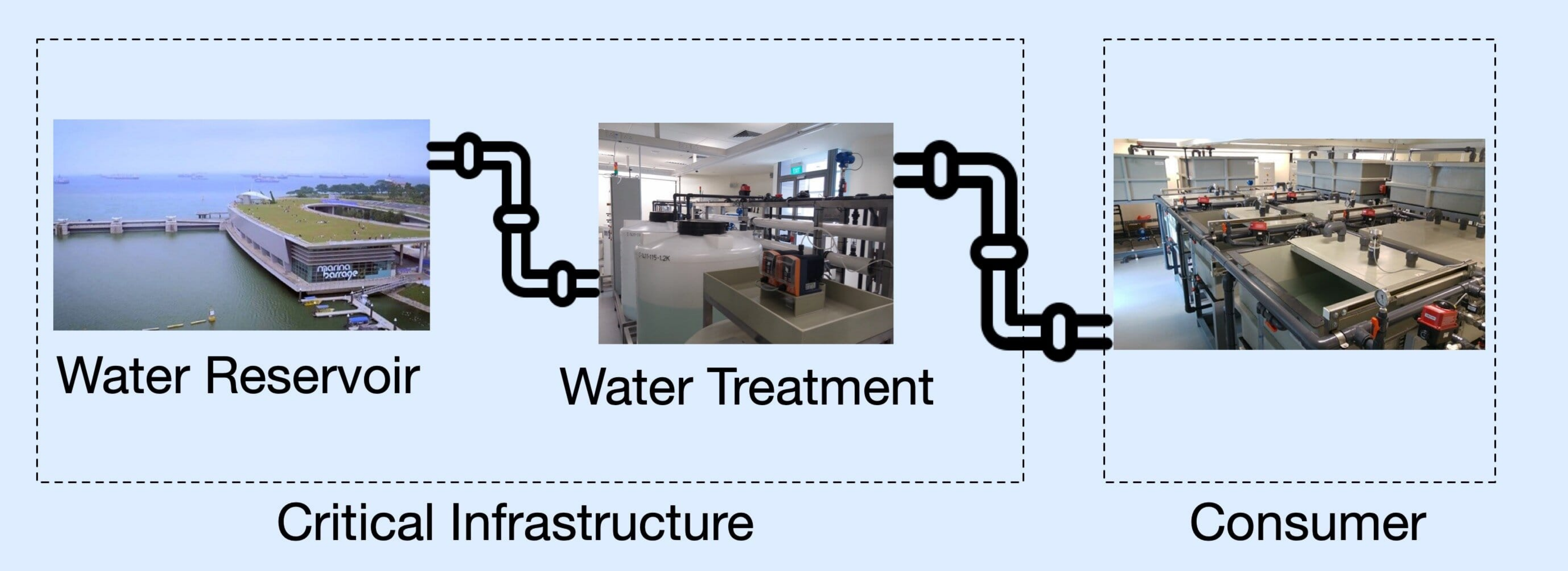}
    \caption{A generic water treatment system as an example of CPS.}
    \label{generic_water_treatment_system_fig}
\end{figure*}

A similar example is a water treatment system which is one of the critical infrastructures of any modern society. Figure~\ref{generic_water_treatment_system_fig} shows a generic overview of a water treatment system, note that the distribution network is intentionally not shown to simplify the illustration in both the power and water systems. Water treatment system employs sensors to measure the flow, pressure, chemical components, level at different nodes, and also equipped with actuators, e.g., motorized valves and pumps to deliver water as required by each consumer. All these processes are controlled and operated autonomously. The automation achieved due to autonomous communication has resulted in efficient monitoring and managing of the physical processes but at the same time opened up these systems for unwanted entities. 

As explained earlier CPS is a broad term and encompasses a lot of interdisciplinary fields. In this article, we focus on industrial CPS similar to examples outlined here. Since a lot of work surveyed in this article is based on an industrial CPS or industrial control system, our proposed device fingerprinting technique is also tested on a water treatment system.  In the following, an abstraction of the well known Purdue architecture~\cite{Williams_purdue_reference_architecture} for each stage of the critical infrastructure is presented.

\begin{figure*}[!htb]
    \centering
    \includegraphics[scale=0.5]{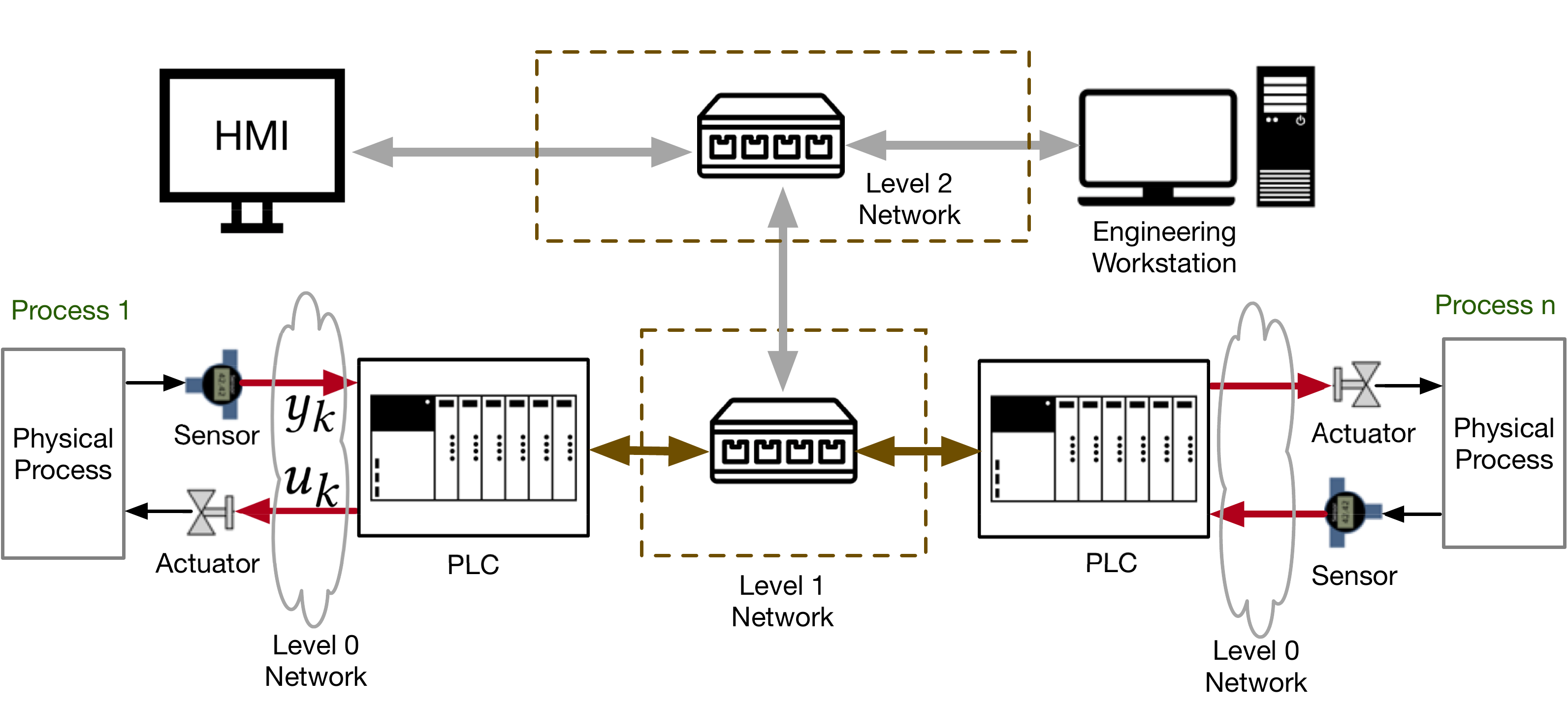}
    \caption[Architecture of a CPS]{An Industrial CPS architecture. Three different communication network levels are shown namely level~0, level~1 and level~2. An attacker can compromise these communication networks as well as the physical components.}
    \label{ics_arch_intro_fig}
\end{figure*}

\subsection{Architecture of an Industrial CPS}

An industrial control system~(ICS) controls a physical process. An ICS takes advantage of the advances in automation technology and interconnected devices. A typical ICS is composed of field devices, e.g., sensors and actuators; control devices, e.g., PLC; monitoring devices, e.g., HMI; control and data logging, e.g., SCADA workstation and programming terminals. In general an ICS follows a layered architecture~\cite{Williams_purdue_reference_architecture}. As shown in  Figure~\ref{ics_arch_intro_fig}, there are three levels of a communication network. Level~0 is the field communication network and is composed of field devices, e.g.,  remote I/O units and communication interfaces to send/receive information to/from PLCs. Using the level~0 network, sensors send the physical process state to the PLCs and in turn, PLCs send the control commands to the actuators. Level~1 is the communication layer used by PLCs to communicate with each other for exchanging data to make control decisions. Level~2 network is used by PLCs to communicate with the SCADA workstation, HMI, historian server; this is known as the supervisory control network.

The communication protocols in an ICS have been proprietary until recently when the focus shifted to using the enterprise network technologies for ease of deployment and scalability, such as the Ethernet and TCP/IP. A survey of communication protocols in an ICS can be found in~\cite{communication_protocol_ICS_survey_2013}. The Figure~\ref{ics_arch_intro_fig}, also represents a specific example of a water treatment testbed used in this study. The communication protocol in the testbed is the Common Industrial Protocol~(CIP). CIP is an application layer protocol on top of Ethernet/IP~(ENIP) to exchange data at level~1 and level~2~\cite{cip_protocol_odva,enip_protocol_rockwell}. The messages between the devices can use either wired media, i.e. IEEE~802.3, Ethernet, or wireless media i.e. IEEE~802.11 WiFi standard. There are two generic types of messages in the CIP/ENIP standard. i.e. explicit messaging and implicit messaging\,\cite{cip_protocol_odva}. Explicit messages use CIP as an application layer protocol and use TCP/IP service to establish a connection. An example is a PLC sending a request message for the exchange of data to another PLC. Implicit messaging, also known as I/O messaging, is used to communicate between PLC and I/O devices. Implicit messages use ENIP protocol on top of UDP/IP service. Implicit messaging is used with time-critical devices, for the reason that those uses UDP and does not need acknowledgment of the transmitted messages as in the case of CIP. Without an authentication mechanism, one could not be sure if these commands are coming from the legitimate PLC.


Input signals to a PLC~($y_k$) can be digital or analog. Digital signals are ON and OFF and analog signals have a continuous range of values. These signals originate from sensors or switches and are represented in the form of voltage or current. For example, a sensor measuring values using 4-20 mA current loop scales the minimum value to 4~mA and the highest value to 20~mA. These analog signals are fed to an analog to digital converter before given to a PLC for processing. Without an authentication mechanism, the integrity of the signals cannot be assured. Similarly, output signals from a PLC are fed to a digital to analog converter before given to the field devices. The output~($u_k$) interface sends control commands to the actuators and also transmits the messages to rest of the PLCs. Without the authentication mechanism, one could not be sure if these commands are coming from the legitimate PLC. 

In the following requirements for a CPS are discussed before mentioning the related security challenges. 
\subsection{CPS Requirements}
CPS monitor and control the physical world and to satisfy the real-world constraints it should be designed to address the following requirements.

\begin{itemize}
    \item \textbf{Real-time Response}: CPS should satisfy the real-time constraints depending on the process. For example, if the process under consideration is electricity the response regarding the sensor measurements should be quick as compared to the water systems. However, each process has its own real-time response constraints which should be fulfilled. Any delays in dissemination of commands due to a fault or an attack (e.g., Denial of Service), can prove to be disastrous. 
    \item \textbf{Resource Constraints}: Most of the devices in a CPS are resource constraint. For example, sensing devices, analog to digital converters, remote input/output~(remote I/O) units and controllers are designed to perform specific functions with the limited memory and processing power. The main idea is for the devices to be robust, function for long time periods e.g., 15-20 years and meet the real-time performance constraints. 
    \item \textbf{Availability}: Shutting down a plant is a much more complicated business than restarting a server. CPS has an important requirement of availability. Critical nature of these systems requires a very high availability as could be the case of temperature regulator in a critical biological process or electric grid. Therefore, upgrading hardware and software is also challenging for CPS due to high up-time. The core idea is to not to interfere with the functionality of the CPS.  
\end{itemize}

\subsection{CPS Security Challenges}
From the above discussion, it is clear that the CPS systems are not the same as the typical IT systems. Both types of systems differ in system requirements and also differ in terms of security requirements/challenges. In general, the security policies for IT systems are defined as CIA paradigm, namely Confidentiality, Integrity and Availability of the data. However, in CPS security the paradigm is the same but in an inverted order by importance, that is, in CPS it is AIC namely Availability, Integrity and Confidentiality. 

\begin{itemize}
    \item \textbf{Availability}: This security property ensures that the system or service is available to the authorized persons. As discussed in the previous section, the availability is the important requirement of a CPS and in terms of security, it is the most important property of the system. Few threats possible are Denial of Service~(DoS) attacks or jamming attacks. 
    \item \textbf{Integrity}: Integrity compromise refers to the modification or destruction of data by unauthorized entities. In CPS an attacker can compromise the integrity of sensor data or the commands transmitted by the PLCs. In IT systems confidentiality is more important than integrity but in a CPS integrity of data is considered more important than to keep it confidential~\cite{cardenas2009challenges}.
    \item \textbf{Confidentiality}: This defines the authorized access to the information. Passwords and data encryption are standard techniques to ensure confidentiality of the data. Although solutions grounded in cryptography, such as those that use TLS, HMACs or other authentication and/or integrity guarantees have been advocated in the context of CPS, historically such countermeasures are not widespread due to limitations in hardware and relative computational cost of such protocols~\cite{John_ACNS2017,cardenas2009challenges}. Since many CPS run legacy hardware and are intended to do so for several years, the problem of raising the bar against authentication attacks by device fingerprinting means is a practical one.
\end{itemize}

\section{Reported Attacks on CPS in Wild}
In this section, few famous CPS attacks are briefly discussed. Following those famous attacks would be a discussion on particular attacks on sensors and PLCs from the academia and industry. 

\paragraph{Maroochy Shire~(2000)} This is an early example of an attack on a CPS executed by a disgruntled employee. The attack was carried out in early 2000 by an employee of a contractor who failed to get a job at Maroochy Shire Council. He used the radio terminals installed by himself to spill the sewage in public parks and streets~\cite{slay_miller_2008}.

\paragraph{Stuxnet~(2010)} This attack is discovered in mid-2010 which targeted Iran's nuclear enrichment facilities~\cite{stuxnet_langner_2011_SnP,stuxnet}. Stuxnet was a highly sophisticated worm which exploited 0day vulnerabilities, relied on root-kits to hide, update itself, used stolen certificates and replayed sensor and network data. It is reported to be a successful attack end up destroying target centrifuges.

\paragraph{Ukrainian Electric Power Grid (2015,2016)} In December 2015 cyberattacks on Ukrainian electric power grid cut off the power supply to customers at the peak of the winter season. The attackers remotely controlled the SCADA distribution system and forced operators to switch to the manual mode which resulted in much longer recovery times\cite{ukranian_case2016analysis}. This attack was over but for another attack to come in the next year around the same time. In 2016 again Ukrainian electric power grid met another cyber attack through the use of Crashoverride malware~\cite{ukranian_case2017analysis_crashoverride}, This attack switched circuit breakers in an unusual open-close pattern in a fast manner, which resulted in cutting off the power supply to the customers.

\paragraph{TRITON Attack~(2017)} This cyber attack was executed on Saudi Arabia's leading oil company Saudi Aramco. The attack was launched using TRITON malware by getting unauthorized access to the engineering workstation. The goal was to reprogram the controllers and cause significant physical damage. This attack forced controllers to enter into a failed safe state disrupting the control of the heavy machinery~\cite{Saudi_aramco_marina_fireeye_2017_triton}.

\paragraph{Norsk Hydro Attack~(2019)} In March 2019 one of the world's biggest aluminum producers Norsk Hydro in Oslo was subjected to a ransomware attack. This attack costed Hydro $\$40$ million in damages~\cite{norsk_hydro_2019_attack_ransomware}.

\paragraph{ASCO Industries Attack~(2019)} This is one of the most recent attacks on CPS. ASCO industries manufacture aerospace parts and got hit by a ransomware attack affecting its production in plants around the world. This attack occurred in mid-June 2019 and the damage is still being assessed~\cite{asco_aircraft_2019_attack_ransomware}.

Few of the famous attacks on CPS are discussed above. In the following specific attacks on the industrial devices are discussed.

\subsection{Sophisticated Attacks on CPS in Research}

An important difference between Cyber Physical Systems (CPS) and traditional IT systems, is that CPS has a physical space to secure besides the cyber domain. In this context, an adversary can also launch an attack from the physical domain, such attacks are not studied in earlier cyber security research. In particular, the \emph{physical} integrity of the CPS, and its availability, are often more important than confidentiality\,\cite{Gollmann2016}.  Moreover, in a CPS an attacker besides compromising the computing elements e.g., sensors through communication networks might also do so from the physical space. This is illustrated for instance by a recent attack\,\cite{drone_Son2015} where a crash is induced in a drone by means of a sound signal that confuses the gyroscope, or by carrying out an analog sensor spoofing attack\,\cite{shoukry2015,yasser-2013,sampling_race2016,ghosttalk_2013}. In~\cite{ghosttalk_2013} attackers would inject data using the sensing device wire as an antenna by intentional electromagnetic interference at the resonant frequencies of the sensing device. In~\cite{walnut_acoustic_attack_mems_accelerometer} a new attack vector is proposed inspired from~\cite{drone_Son2015}. A modulated audio signal could result in desired data injection~\cite{walnut_acoustic_attack_mems_accelerometer}. A recent study has shown sonic attacks for a range of smart sensing devices~\cite{sonic_gun_blackhat2017}. Anti-lock braking system~(ABS) is attacked in real vehicles using the signal injection in the analog/physical domain~\cite{yasser-2013}. A recent article~\cite{trickorheat_kevinfu_temp_sensor_attacks_EMI_2019} attacked temperature sensor in infant incubators using electromagnetic signals.    Thus, security requirements for CPS introduce new challenges and hence the need to expand traditional attacker models to include physical and cyber-physical characteristics of a system\,\cite{marco_cpdy2016}, and consequently introduce a need for novel security solutions.

\subsection{Attacks on PLCs}

Guaranteeing data integrity in the presence of strong adversaries, for instance 
against those who can gain full control over PLCs, is challenging. For instance, a study reported in~\cite{eireann_2013_greyhat_PLC_vulnerabiity} reveals that a large number of PLCs are connected to the Internet and contain vulnerabilities related to authentication. Using the discovered vulnerability, the authentication mechanism is bypassed and full control over the PLC could be achieved over the internet. The use of commercial off the shelf~(COTS) devices in a CPS, and software backdoor, can lead to full control over PLCs~\cite{ruben_backdoors_PLC_2012_blackhat}. In~\cite{fovino_2009_malware_PLC} authors have used lack of authentication in the Modbus protocol to take over the controllers and send unauthorized commands to the other devices. 
Stuxnet is a famous example of a malware attack where PLCs were hijacked and malicious code altered the PLC's configuration~\cite{stuxnet_langner_2011_SnP}. Attackers have executed web-based DoS and resetting PLC attacks by exploiting bugs in PLC code which were connected to the internet~\cite{robert_turk_2005_PLC_DoS}. Recently a range of malware and network-based attacks were designed and executed against PLCs~\cite{Anand_ESORICS2017_PLC_ladderlogicbomb,MS_thesis_PLC_attack_NTNU_2013}. Therefore, there is a need for authenticating CPS devices non-invasively and without disturbing their core functionality.


\begin{figure}[!htb]
\centering
\includegraphics[width=3in,height=2in]{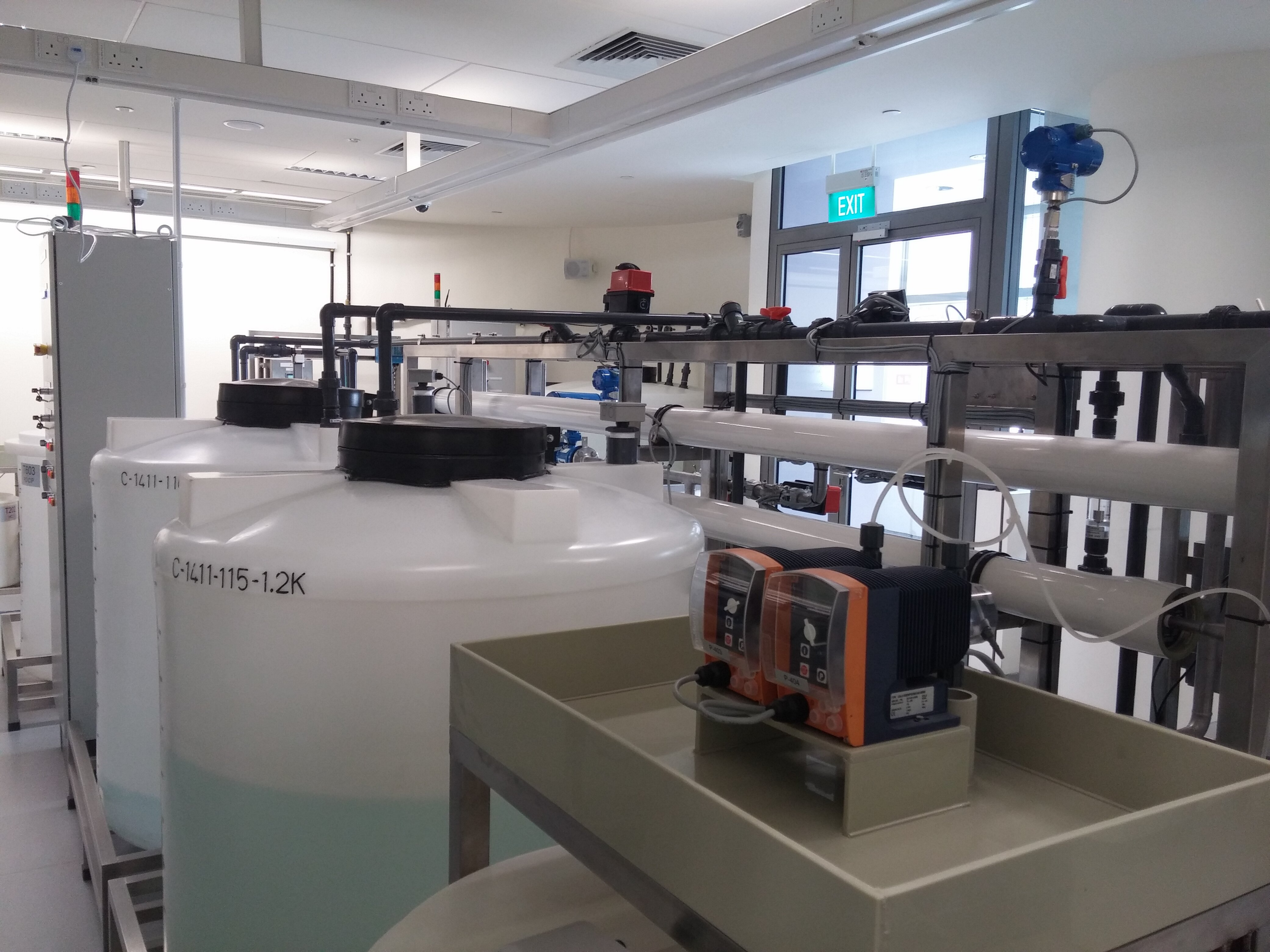}
\caption{Experiment Setup: Secure Water Treatment Testbed Plant Layout (SWaT)}\label{fig_swat_testbed_photo}
\end{figure}

\begin{figure}
    \centering
    \includegraphics[scale=0.35]{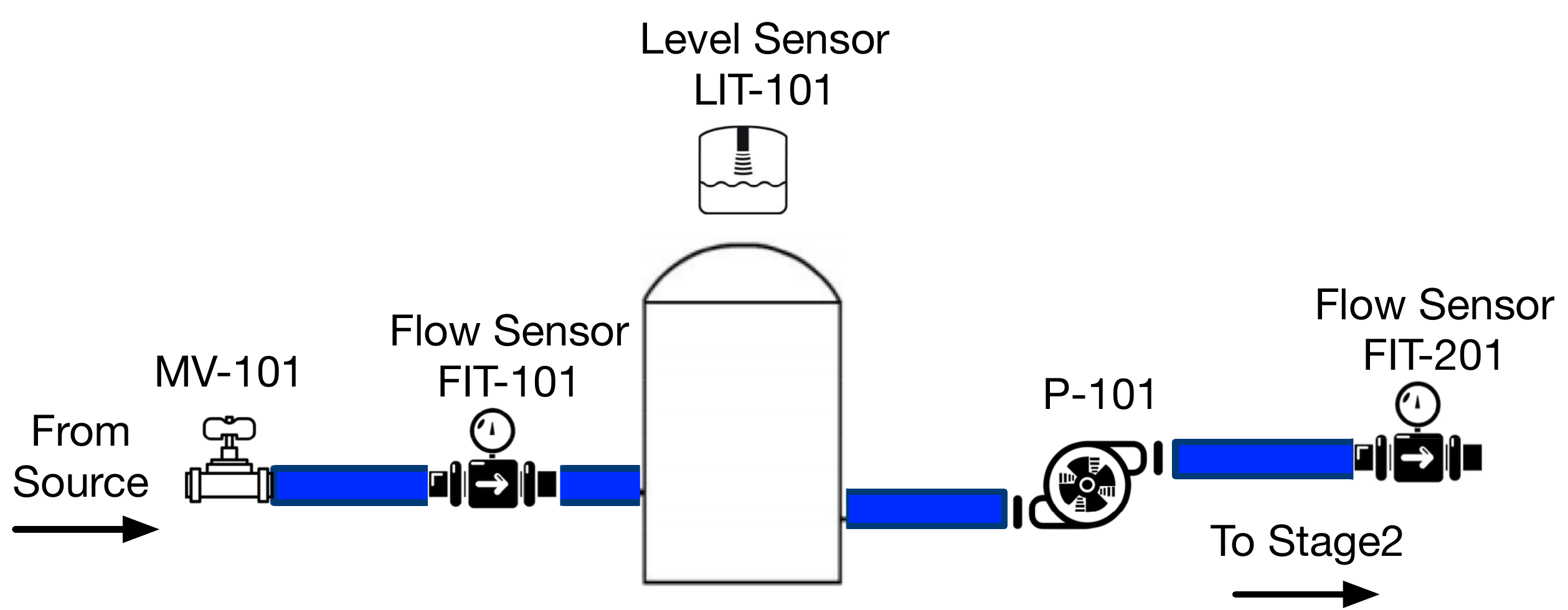}
    \caption{A partial setup from a water treatment plant as a motivating example.}
    \label{fig:tank1_modeling}
\end{figure}

\section{Physics based Perspective}

\subsection{A Motivating Example}


We will present our findings based on experimentation done in a water treatment plant. Figure~\ref{fig_swat_testbed_photo} shows a picture of the testbed used. It is a six-stage water treatment process, for details refer to the testbed paper~\cite{swat2016}.
We will use the first stage of the water treatment process as a motivating example. A physical system diagram for stage~1 is shown in Figure~\ref{fig:tank1_modeling}. 
Figure~\ref{fig:tank1_modeling} shows a level sensor mounted on top of the water tank to the water level and the inflow and outflow of the water is being controlled by the motorized valve (MV-101) at the input and pump (P-101) at the output respectively. The idea is to model this inflow and outflow by considering the physical principles and the design of the physical process. For a tank, we know that the rate of change of water inside the tank is equal to the difference between water flowing into the tank and water flowing out from the tank with respect to time. We can represent this using the mass-balance equation such as,  
\[ \frac{dV}{dt} = Q_{in} - Q_{out} \] 
\begin{equation}\label{mass_flow_eq}
\frac{dh}{dt} = \frac{Q_{in} - Q_{out}}{A} \ \ \textrm{since} \ \ V=A\times h,
\end{equation}
where $V$ represents the volume of the tank, $A$ is the cross-sectional area of the tank, and $h$ is the height of the water inside the tank, \eqref{mass_flow_eq} provides a linear equation, we can see the term $[Q_{in}- Q_{out}]$ represents the water flow which depends upon the PLC control actions implemented via MV-101 and P-101. From Figure~\ref{fig:tank1_modeling}, it can be seen that using the height and diameter of the tank from design documents, it is possible to figure out the volume and the cross-sectional area of the tank. Let us consider that state of the physical process as the height of water inside the tank. Then the solution of this equation gives us the following result.
\[x_{k+1} = x_k + u_k,\]
where $u_k$ is the PLC control action. Here $x_k$ represents water level in the tank at time $k$. The control action $u_k$ can be a either open/close (for the motorized valve) or on/off (for the pump). Similarly we can describe the sensor state and we can get the set of system equations. Following represents the systme dynamics in form of a state space model.

\begin{equation}\label{system_eq}
\begin{cases}
x_{k+1} = Ax_k+ Bu_k + v_k, \\
y_k = Cx_k + \eta_k.
\end{cases}
\end{equation}

Where $y_k$ is the sensor measurement driven by the control action $u_k$. Matrices $A, B$  and $C$ are the state-space matrices of appropriate dimensions. $v_k$ and $\eta_k$ are the process and measurement noise vectors respectively. From \eqref{system_eq}, it can be seen that if we have a system state value at time $k$, then given the PLC control $u_k$ we can predict the next state at time $k+1$. 
For example, if the MV-101 control is set to open the valve and P-101 as turned ON, given the information of this control from PLC, we know from the design of the physical process that how much the water level in the tank should increase. This is an example of how can we use the physics of the system to model the physical process. Once the system model has been obtained it is possible to learn the normal behavior of the process in a mathematical form.

\begin{figure}[t]
\centering
\includegraphics[scale=0.30] {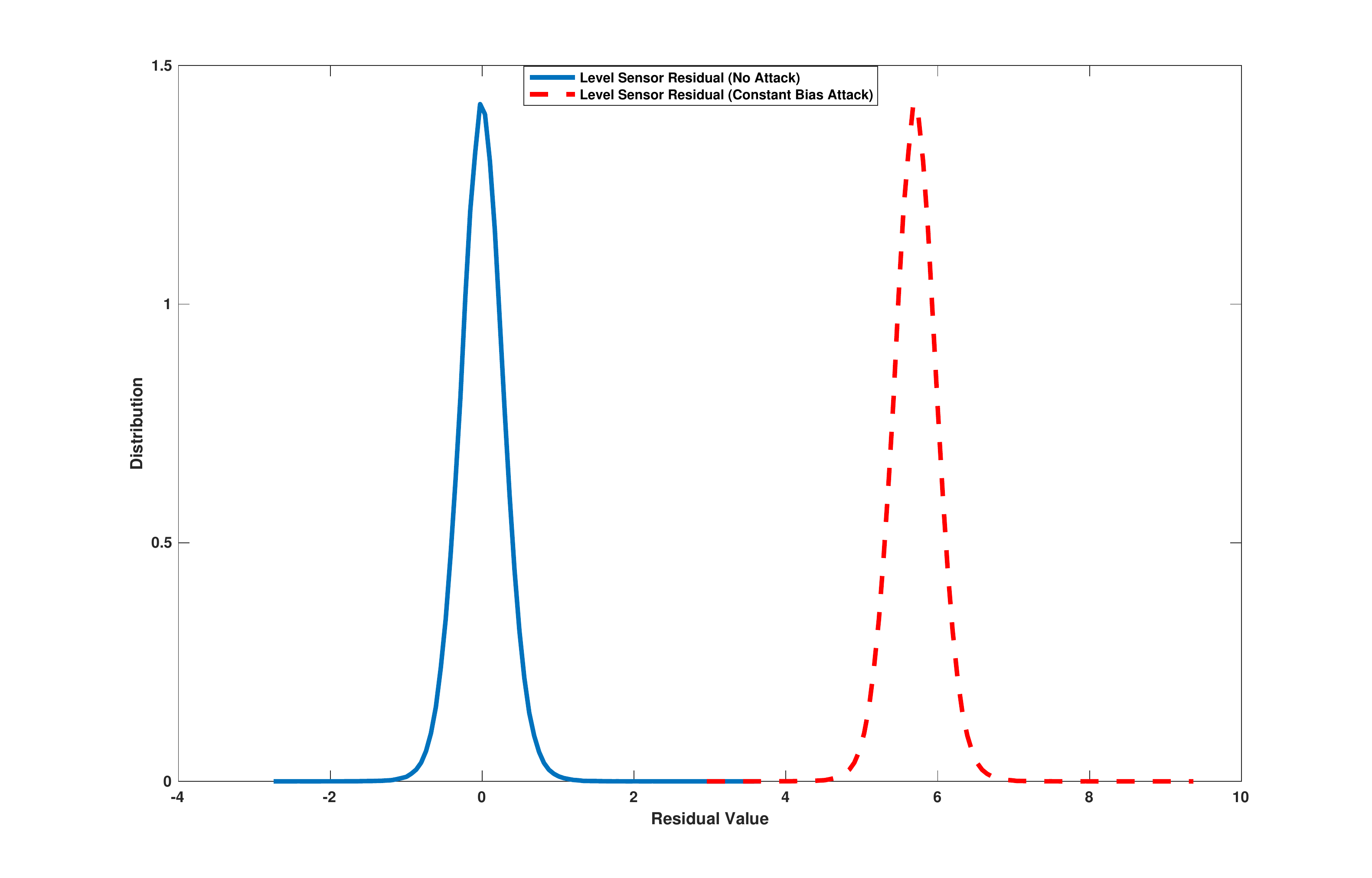}
\caption{(Left): Probability distribution of the residue for level sensor measurements without attack. (Right): Probability distribution of the residual for water level sensor measurements with bias injection attack.}
\label{pdf_residue}
\end{figure}

\subsection{Attack Detection Framework}
A general attack detection framework has two major components, 1) system model and estimation and 2) a threshold based detector.

\Paragraph{System Model and Estimation}
The idea of obtaining a system model is explained in the previous section. The system models can be obtained either using data based techniques or the first principles~\cite{Rizwan_ESORICS2017_stealthyAtt,Carlos_Justin_CDC2016_stealthyAtt,urbina_CCS2016limiting,pasqualetti2013attack_control4}. Using the system model it is possible to estimate the states of the system and ultimately estimate output from a sensor~($\hat{y}_k$). A residual vector is calculated by taking the difference between the sensor measurements and estimated sensor output as,

\begin{equation}
    r_k = y_k - \hat{y}_k.
\end{equation}

\noindent Where $r_k$ is the residual vector. For the residual, the hypothesis testing is for $\mathcal{H}_0$, the \emph{normal mode} (no attacks), and $\mathcal{H}_1$, the \emph{faulty mode} (with attacks). The residuals are obtained using this data along with the state estimates. Thus, the two hypotheses are stated as follows,
\begin{center}
\begin{tabular}{ c c c }
 $\mathcal{H}_0: \left\{
\begin{array}{ll}
E[r_k] = 0,  \label{29} \\[.5mm]
E[r_kr_k^T] = \Sigma, 
\end{array} \text{or}
\right.$ & $\mathcal{H}_1: \left\{
\begin{array}{ll}
E[r_k] \neq 0,\   \label{30} \\[.5mm]
E[r_kr_k^T] \neq \Sigma.
\end{array}
\right.$
\end{tabular}
\end{center}

\Paragraph{Threshold based Detector} To detect the presence of an attack, the residual vector is tested against a predefined threshold designed for a particular false alarm rate. Figure~\ref{pdf_residue} shows the distribution for a residual vector with a mean value of 0 without an attack and the second plot in case of an attack. We can create a threshold for the residual distribution and if the values of residual are outside that threshold declare it under an attack,

\begin{equation}\label{threshold_detector_eq}
    |r_k| > \tau, Alarm = True.
\end{equation}

\noindent Where $\tau$ is a threshold and |$r_k$| is the absolute value of the residual. There have been studies on optimizing the parameters of different stateful and stateless detectors~\cite{carlos_msc2016,urbina_CCS2016limiting}. A wide variety of algorithms exist to chose the best threshold value to maximize the attack detection rate and minimize the false alarm rate. 

\subsection{Prior Research}

In this section, we will highlight the research that has been done in CPS security exploiting the physical models of the process. A general approach is to, 1) create models of the normal process either based on the data from simulations/real systems or based on the first principles, and 2) use the statistical detectors to find if there are any deviations from the normal/expected behavior. 

One of the earlier works on the security of power systems against data injection attacks is detailed in \cite{liu_ccs2009_first_work_}. Authors had shown that a bad data detector would raise an alarm for random attacks similar to a fault but not a stealthy attack. Another study related to a smart water distribution system \cite{Ahmed_AsiaCCS2017_stealthyAtt} has also made similar observations. These studies created the models of the physical process based on simulations and the real testbed respectively. 

\Paragraph{Process/Physical Invariants} The idea of invariants is to model the physical states as such that certain physical laws shall be obeyed. Invariants are designed using the relationship between different state variables. No matter what happens these relationships should not vary. Designed invariants are using physical laws underneath to ensure the laws of physics are being obeyed. A state relation based intrusion detection is proposed in \cite{wang2014srid_ESORICS2014_invariants}. This study used a relational graph to model the different nodes related to each other via a physical principle. Similar research is conducted on a water treatment system ~\cite{adepuMathurIFIPSEC2016_invariants} by creating invariants from the physical process. A more recent effort on similar lines is to create control invariants ~\cite{choi2018detecting_dongyan_ccs2018_drone_control_invariant}. The authors tested their approach on a drone.

\Paragraph{Active Defense} Some techniques use active methods to detect attacks. These techniques are a combination of modeling the physics of the system and active detection methods. A challenge-response based sensor attack detection technique is presented in \cite{shoukry2015}. The proposed technique is tested on vehicles for active sensors.  Another active technique called as physical watermarking is proposed in \cite{bruno2015_watermarking}. 

\Paragraph{Control Theory/State Estimation} Most of the physics-based detection techniques originate in control theory due to a history of literature on modeling the physical processes. Also, fault detection in control systems has been studied extensively over the past half-century. There are several works on using the model of a physical process \cite{bai2014kalman_control1,CPSweek2016_stealthy_replayATT,Carlos_Justin_CDC2016_stealthyAtt,pasqualetti2013attack_control4}. Most of these works borrow ideas from fault detection literature and has also contributed towards the limitations of fault detectors to be used as attack detectors. Towards that end, secure state estimation has extensively been studied. Recently,  a research work in~\cite{shoukry2018smt_control3}
proposed a search algorithm based on Satisfiability Modulo Theory (SMT) to speed up the search
of possible sensors sets, followed by an extended work to model the noisy systems~\cite{mishra2016secure_estimation_control2}.

\Paragraph{Unsupervised Learning} The problem with a supervised learning detection method is that it needs to learn the normal model as well as from the data under attack. In real-world availability of attack data is a big issue, therefore, some studies employ semi-supervised or unsupervised learning for attack detection. In the following a couple of the recent works~\cite{noisematters_Mujeeb_acsac2018,krotofilLarsenGollman_process_matters} are discussed, those used the model from the plant dynamics and unsupervised learning for attack detection. A signal entropy based detector is used in \cite{krotofilLarsenGollman_process_matters} and one-class SVM is used in \cite{noisematters_Mujeeb_acsac2018} as a detector. 

\Paragraph{Physical Authentication} There have been some interesting efforts to authenticate the control logic in a PLC by using the physics of the process~\cite{roth2016physical_authentication_BruceMcmillan,sunjun_ieee_snp_2018_dataset_swat}. One recent study had exploited the physics of the process to discover an insider threat~\cite{Anand_agrawal2018poster_AsiaCCS2018}.

\Paragraph{Evaluation Metrics} A recent work in \cite{urbina_CCS2016limiting} proposed a new evaluation metric for the physics based attack detection algorithms. They considered a case of a stealthy attack and measured its impact on the physical process. The list here is by no means exhaustive, the intention is to give readers an idea of how popular physics based methods are in the CPS security.

\subsection{Shortcomings of Prior Works}


\Paragraph{Interfering Techniques} Active defense techniques, for example, watermarking or challenge-response can be considered interfering with the normal operation of the process. In the case of physical watermarking techniques, a noise signal is added to the optimal control signal which can degrade the performance of the system under study. Similarly in challenge-response techniques, a challenge affects the performance of the active sensors due to the introduced challenges. For a CPS a non-interfering passive technique would be preferred.

\Paragraph{Number of Devices under Attack} State estimation based and invariants based techniques rely on the relationship between sensors and actuators. If all the sensors and actuators are under attack then model based methods shall fail. Therefore, it is desired to design a technique that can identify attacks on devices independently from other devices.

\Paragraph{Stealthy Attacks} Most of the work using a system model along with a statistical detector is prone to a smart attacker. For example, if an attacker learns a threshold for the statistical detector and stays below that, it does not get detected. From Eq.~\eqref{threshold_detector_eq} we can rewrite the expression as,

\begin{equation}
    |y_k - \hat{y}_k| > \tau.
\end{equation}

\noindent If sensor measurements are under attack~($\delta_k$) then the attacked sensor measurement goes to $y^a_k = y_k + \delta_k$ resulting in,

\begin{equation}\label{stealthy_eqn_1}
        |y_k + \delta_k - \hat{y}_k| > \tau.
\end{equation}

\noindent Remember $\delta_k$ is the attacker's signal and it can choose it to be anything. An attacker can always choose $\delta_k = \hat{y}_k - y_k + \tau$ which will change the expression in Eq.~\eqref{stealthy_eqn_1} to $\tau > \tau$, which is never true and no alarms will be raised although attacker is injection a value of $\tau$ at each time step in the sensor measurement. Such an attack is considered to be stealthy and in theory, can be designed for any threshold based detector. 



\Paragraph{Lack of Testbed-based Validation} Most of the previous studies are performed either on a dataset or a simulation based model. It is important to validate the proposed techniques on real systems or testbeds to identify challenges which an operator or plant engineer might face when the system is under attack or due to false alarms.

\subsection{Suggested Improvements}

\Paragraph{Passive Attack Detection Techniques} Given the critical nature of the industrial systems, it is desired to have a passive technique for attack detection. We can not afford legacy ideas of active defense from the IT security literature.

\Paragraph{Assumption on Number of Devices under Attack} Ideally the proposed attack detection techniques shall be independent of the number of devices under attack. We should come up with the methods where it would be possible to identify attacks on each device separately. 


\Paragraph{Validation on Testbeds or Real Systems} Most of the previous studies are based on the simulations. It is easier to work with simulation models but those studies miss details that are encountered in practice by industrial engineers. 

\noindent In the following, we will summarize ideas related to authenticating devices based on the hardware characteristics of the devices, passively. 

\begin{figure*}
    \centering
    \includegraphics[scale=0.4]{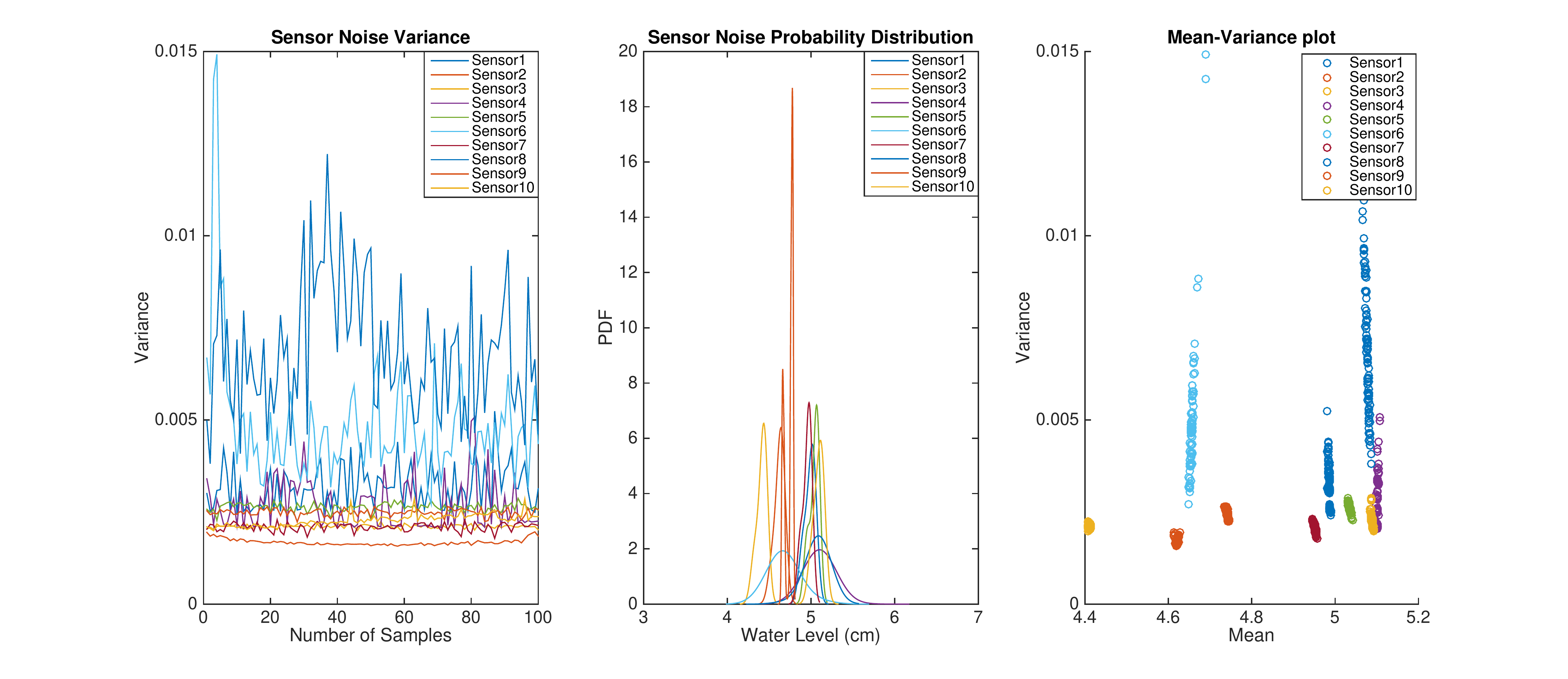}
    \caption{Sensor noise from 10 ultrasonic level sensors and their noise vector distribution.}
    \label{variance_plot}
\end{figure*}

\begin{figure*}
    \centering
    \includegraphics[scale=0.6]{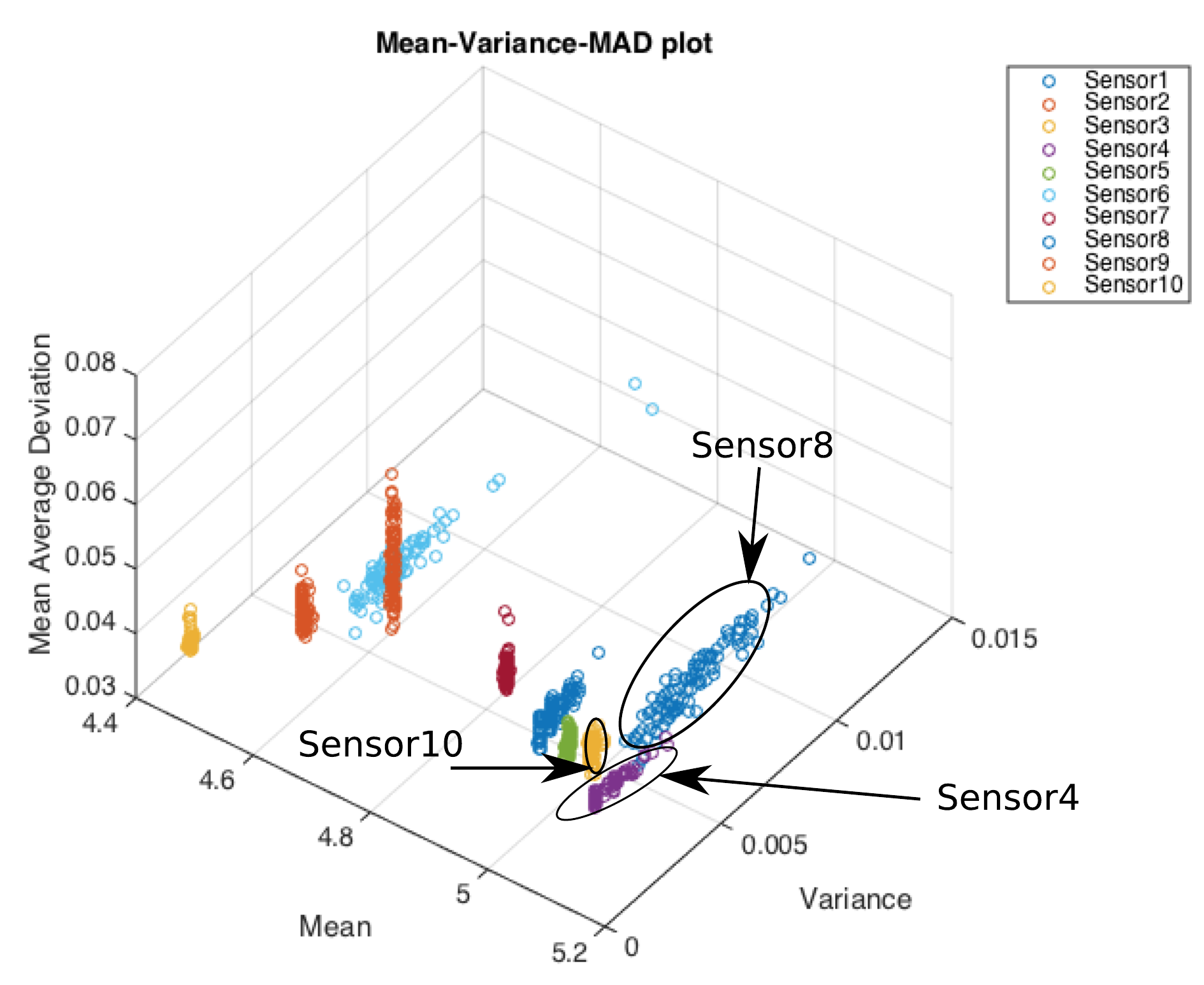}
    \caption{A proof of existence of noise based sensor fingerprint for all the level sensors of same type and model based on three time domain features.}
    \label{mad}
\end{figure*}

\section{Device Fingerprinting}


A device fingerprint refers to some unique features of a device's hardware, software or a combination of both. Device fingerprinting ideas have been tested in different domains. The idea of fingerprinting a PC remotely based on its clock skew is presented in\,\cite{kohno2005}. Small microscopic deviations in device's clock\,\cite{moon1999,paxson1998} is  used as a fingerprint. In\,\cite{raheem2015} inter-arrival time of packets is analyzed to fingerprint devices on a small campus network. In\,\cite{dey-2014} hardware imperfections during the sensor manufacturing process are exploited as a fingerprint for a smartphone. In CPS a recent work tried to create fingerprints for the actuators based on the opening/closing times~\cite{raheem2016}. Device fingerprinting techniques to authenticate devices and passively detect attacks have been found promising in the IT domain but for industrial-grade sensors, such a study was needed. In the following, we briefly discuss one of our proposed technique titled \acronym~\cite{NoiSense_eprint_Mujeeb}.

{\acronym} is proposed as a non-intrusive sensor fingerprinting technique to authenticate sensors transmitting measurements to one or more PLCs. 
Device fingerprinting ideas existed in other fields as mentioned above, however, sensors in a CPS are not functionally/computationally similar enough to exhibit the above-mentioned fingerprints~\cite{raheem2016}. Thus, we seek an answer to the question, \textit{Do sensors in a real-world CPS have unique fingerprints?} It is known that hardware imperfections during the manufacturing process exhibit some unique physical behaviors that are useful for profiling and fingerprinting~\cite{dey-2014}. In particular, we observe that \emph{noise} (imperfections in measurements), an otherwise undesirable feature of sensors, strongly depends on such manufacturing imperfections. These variations affect each device differently and thus are hard to control or reproduce~\cite{gerdes2006}, making it challenging for an attacker to imitate sensor noise patterns. 

\acronym creates a fingerprint for a sensor based on a set of time domain and frequency domain features that are extracted from the sensor noise. A machine learning algorithm is used to distinguish an individual sensor from others. 
Experiments were performed on sensors of different types in an operational water treatment and distribution facility accessible for research\,\cite{swat2016,wadi2017}. Sensor identification accuracy is observed to be as high as $97\%$, with a low of $90\%$. It is also shown that the proposed scheme is scalable for tens of sensors and that the sensor fingerprint is stable over time. The true positive rate for sensor identification is observed to be  $100\%$ for most of the sensors and false positive rate as low as $0\%$, see \cite{NoiSense_eprint_Mujeeb} for details. 

\noindent \emph{Does a unique fingerprint exist for each sensor?} A  limited number of sensors were available in the water utility testbeds. Hence,    additional low-cost ultrasonic sensors are included to explore the existence of fingerprints for many sensors of the same type and model. To demonstrate the existence of fingerprint,  ten dual transducer ultrasonic sensors (HCSR04) from the same manufacturer were used. All ten sensors were mounted on the same water tank. Data was collected for 3~hours and many chunks of the collected data taken for analysis. Each chunk consists of 300 readings from the sensor. Figure~\ref{variance_plot} shows results for the collected data. The plot on the left shows the variance of noise vector from each sensor for all chunks. It is observed that some of these sensors have a unique noise variance and can be distinguished from each other but there remain few sensors that have similar noise patterns in terms of noise variance. The middle pane is a plot of the distribution of the noise vector from each sensor. It also shows that sensors can be distinguished based on noise statistics. However, there remain overlaps among some sensors. The right pane shows 2-D clustering of the sensors. Sensors can be distinguished more precisely by using one more feature of sensor's noise i.e. mean value. The scatter plot on the right-hand side clusters each chunk with its respective mean and variance. The separation is quite clear but there remain  overlaps, e.g.,  sensor4, sensor8 and sensor10. We need additional features to further eliminate such overlaps. In Figure\,\ref{mad}, by adding one more feature, i.e.  mean average deviation, sensor4, sensor8 and sensor10 can be distinguished. 

\section{Sensing Technologies and Basis for Fingerprints}\label{sensing_technologies_appendix}
In this section, we explain the working principle of the sensing technologies under study. This insight in sensor construction and functionality is an aid in understanding the sources of sensor noise and fingerprints.

\subsection{Ultrasonic Level Sensors}\label{ultrasonic-technology}
Water treatment testbed uses ultrasonic sensors based on a piezoelectric (PZT ceramic) material transducer. The level of water in a tank is calculated by measuring the return time of the acoustic wave after hitting the water surface.  Several factors  contribute to variations in the measurements obtained from  ultrasonic sensors. These  measurements depend on the speed of sound which changes according to the surrounding temperature. Speed of sound through air as a function of temperature can be expressed as\,\cite{jenny-2013},

\begin{equation} \label{speed_of_sound}
   c_{air}(t) = C_{0} + Kt,
\end{equation}

\skipnoindent where, $t$ is the temperature in degree Celsius; $K$ is the rate of change of speed, which is approximately 0.607 m/s at every 1 degree Celsius change; and $C_{0}$ is the speed of sound in air at 0 degree Celsius which is 331.45 m/s. Besides temperature, obstacles like tank walls reflect echo sooner than it should be, contributing towards noise in the measurements. Water sloshing is another reason for erroneous level measurements. Ultrasonic level sensors depend on PZT ceramic transducer to convert sound waves into electrical signals. These PZT materials convert sound vibrations to an electric signal. The acoustic impedance of these transducers also depends on temperature thus adding another source of noise \cite{coutard-2005}. Thermal and polarisation noise are the main sources of voltage fluctuation in piezoelectric ceramics. Thermal noise originates from interaction of phonons with free electrons or holes. The spectral density of this noise is proportional to sensor resistance and temperature. Electrical polarization in piezoelectric materials is also a source of voltage fluctuation\,\cite{petr-2011}.    


 \begin{figure}[!htb]
\centering
\includegraphics[scale=.2]{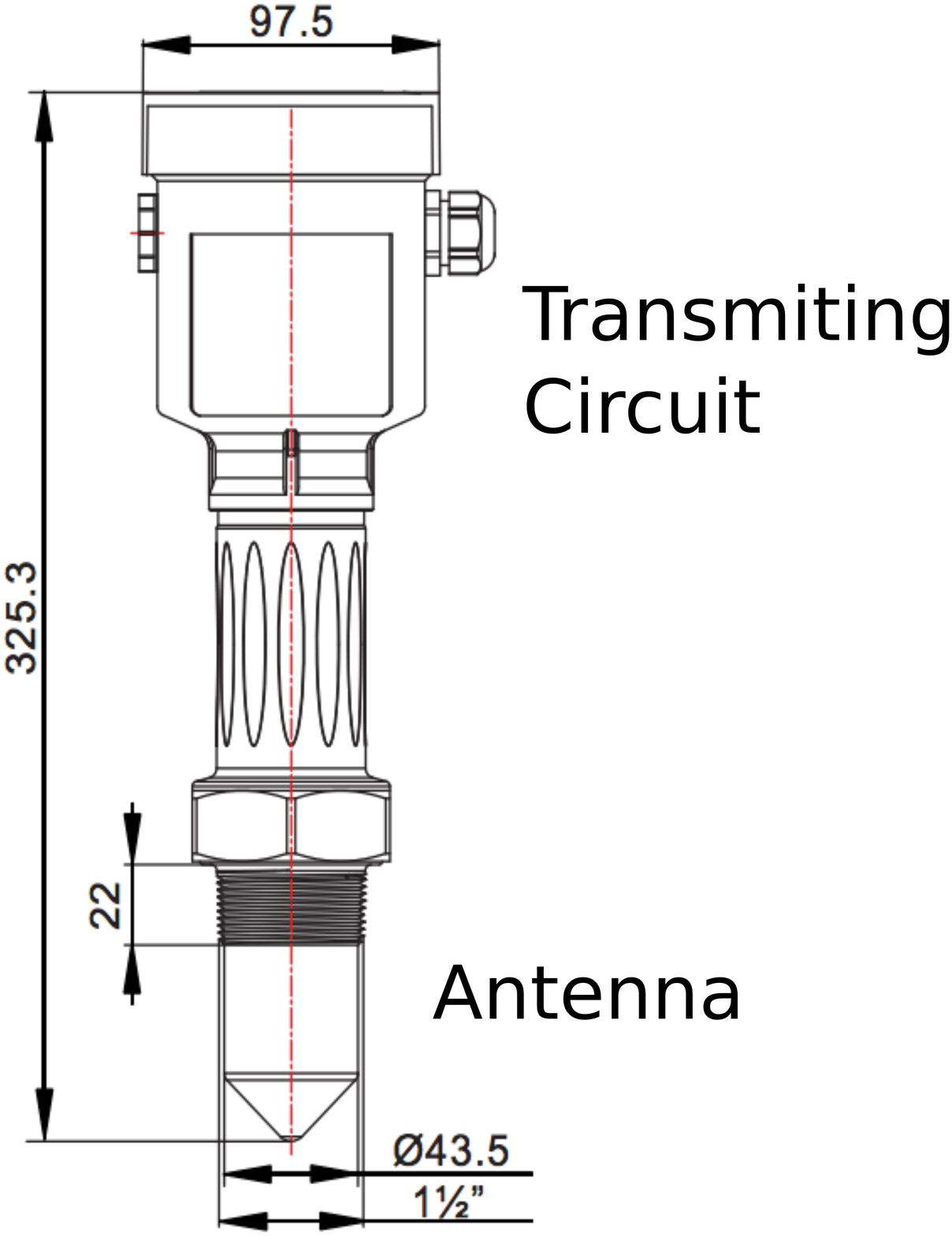}
\caption{RADAR level sensor construction. Antenna is the element responsible to capture microwaves reflected from the water surface. Operating frequency is 26 GHz\,\cite{flotech_radar}.}
\label{radar_levelsensor}
\end{figure} 


 \begin{figure}[!htb]
\centering
\includegraphics[scale=.4]{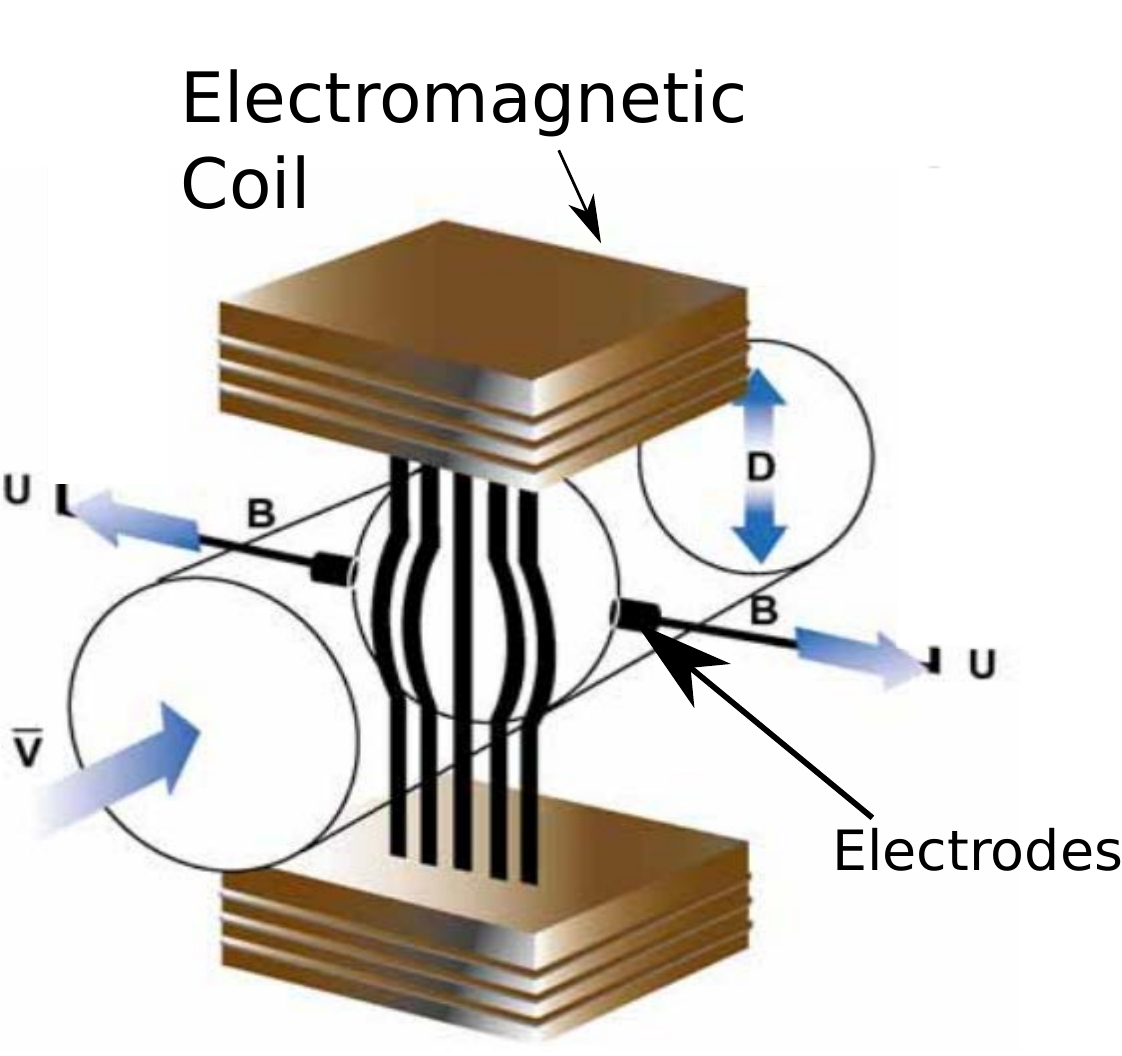}
\caption{Electromagnetic flow meter structure. Electromagnetic coils generate a constant electric field. When water (conducting fluid) flows through magnetic field, a voltage proportional to water speed, is induced at electrodes~\cite{flotech_flowmeter}.}
\label{flowmeter}
\end{figure}  

\subsection{Microwave Level Sensors}
The microwave level/distance sensor, often called RADAR (RAdio Distance and Ranging) works in a similar way as ultrasonic sensors. A microwave pulse is emitted by the antenna that travels at the speed of light and upon hitting the surface of the target it is reflected back and received at the same antenna. 
The distance between the antenna and target is calculated based on the time it takes for the microwave to travel that distance. 
In the  case study reported here the waves are bounced back by water with a dielectric constant of 80 (stronger reflections) which is higher than dielectric constant 1 (no reflection) of free space. This implies that   enough energy will be reflected and reach the antenna. 

Figure~\ref{radar_levelsensor}, shows the microwave level sensor used in water distribution testbed\,\cite{wadi2017}. Similar to an ultrasonic sensor where the sound wave hits the transducer to produce an output voltage and calculate the distance, in microwave based level sensor it is the antenna where the electromagnetic energy is received and distance  calculated. These antennae are designed to have a 50$\Omega$ resistance so that once connected with a cable of characteristic impedance of 50$\Omega$, maximum power transfer takes place from the antenna. The sensor under consideration is designed to operate at 26\,GHz with a beam angle of $22^o$ and 1$\mu$W effective radiated power\,\cite{flotech_radar}. However, in practice these specifications have deviation for the same type and design of an antenna due to manufacturing imperfections and installation inaccuracies. For example, antenna connection with a cable will result in impedance variations\,\cite{accuracy_radar}. Also, beam angle and radiation pattern  varies for each antenna leading to deviations from theoretical design resulting in different range resolution that is ultimately reflected in sensor noise \cite{antenna_pattern}.  



\subsection{Electromagnetic Flow Meters}
The electromagnetic flow meters follow Faraday's law of induction according to which a voltage is induced by an electrically conductive fluid passing through a magnetic field. In an electromagnetic flow meter, the medium acts as the electrical conductor when flowing through the meter tube, and the induced voltage is proportional to the average flow velocity (the faster the flow rate, the higher the voltage). The induced voltage is picked up by a pair of electrodes, mounted in the meter tube, and transmitted to a flow transmitter to produce various standardized output signals. Using the pipe cross-sectional area, the flow volume is calculated by the transmitter. The following equation is applicable to the induced voltage:

\begin{equation}
U = K * B * V * D
\end{equation}

\noindent where $U$ is the induced voltage, $K$ is the instrument constant, $B$ is the magnetic field strength, $V$ is the mean velocity of the fluid, and $D$ is the pipe cross-section.

A commercial electromagnetic flow meter is shown in figure~\ref{flowmeter} \cite{flotech_flowmeter}. 
It's internal structure consists of a pair of coils mounted on the top and bottom of an electrically insulated flow tube. A pair of electrodes protrude through the flow
tube wall perpendicular to the pipe axes and largely normal to the direction of the generated magnetic field. As the liquid passes through the pipe, it moves through the magnetic field and the positive and negative ions within the liquid
experience a force upon them. The forces on the ions cause them to migrate and result in an electric field being generated across the pipe. The Voltage generated across the pipe is measured between the electrodes. 
Noise in these sensor readings come from the area of the electrodes and size of the electro-magnets generating electromagnetic field $B$. The installation and alignment of electrodes and coils will result in different stray capacitance and noise \cite{flowmeter2006}.

     
     \begin{figure}
         \includegraphics[scale=0.55]{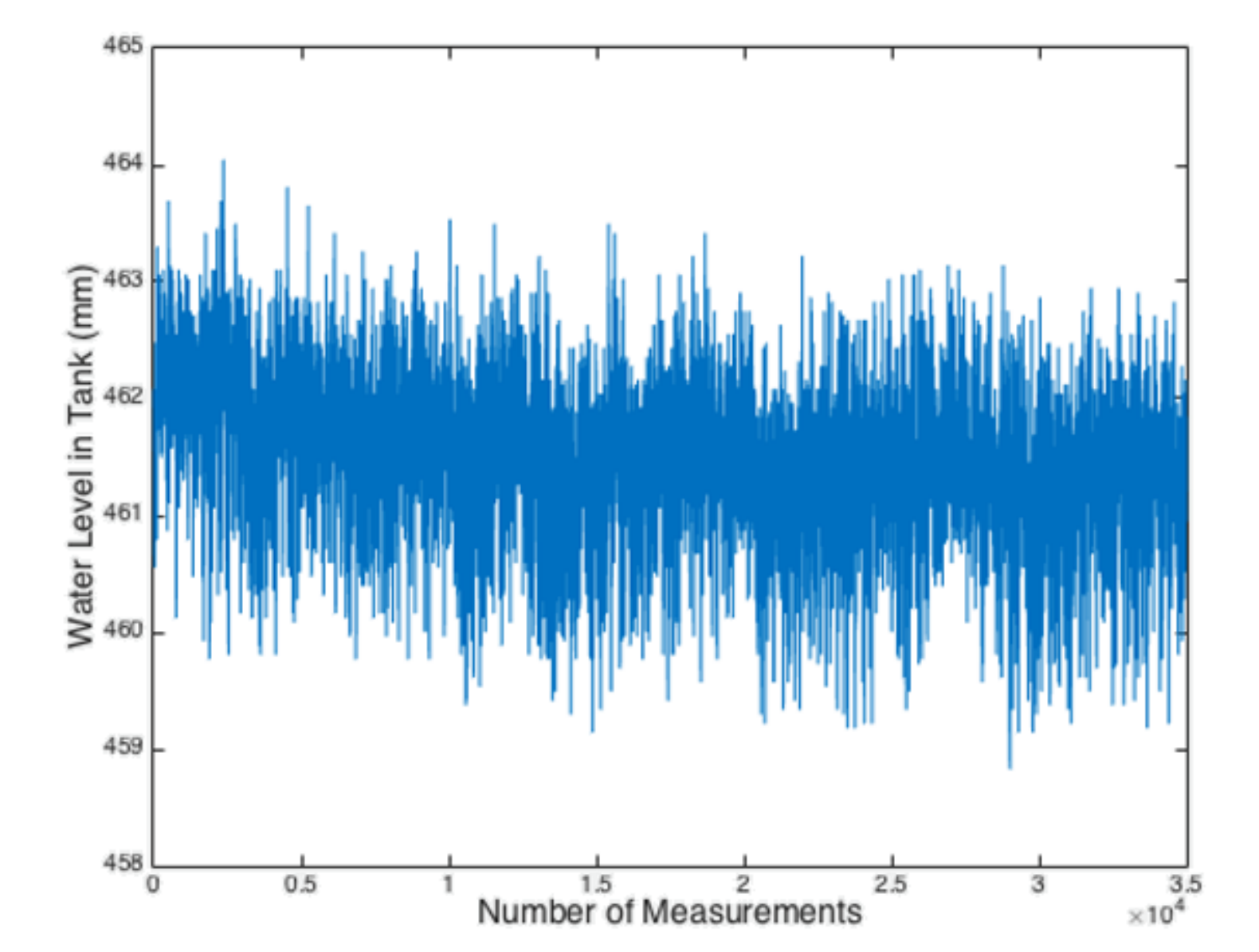}
         \caption{Time series data from a level sensor for a constant water level.}
         \label{fig:time_series}
     \end{figure}

     \begin{figure}
         \includegraphics[scale=0.55]{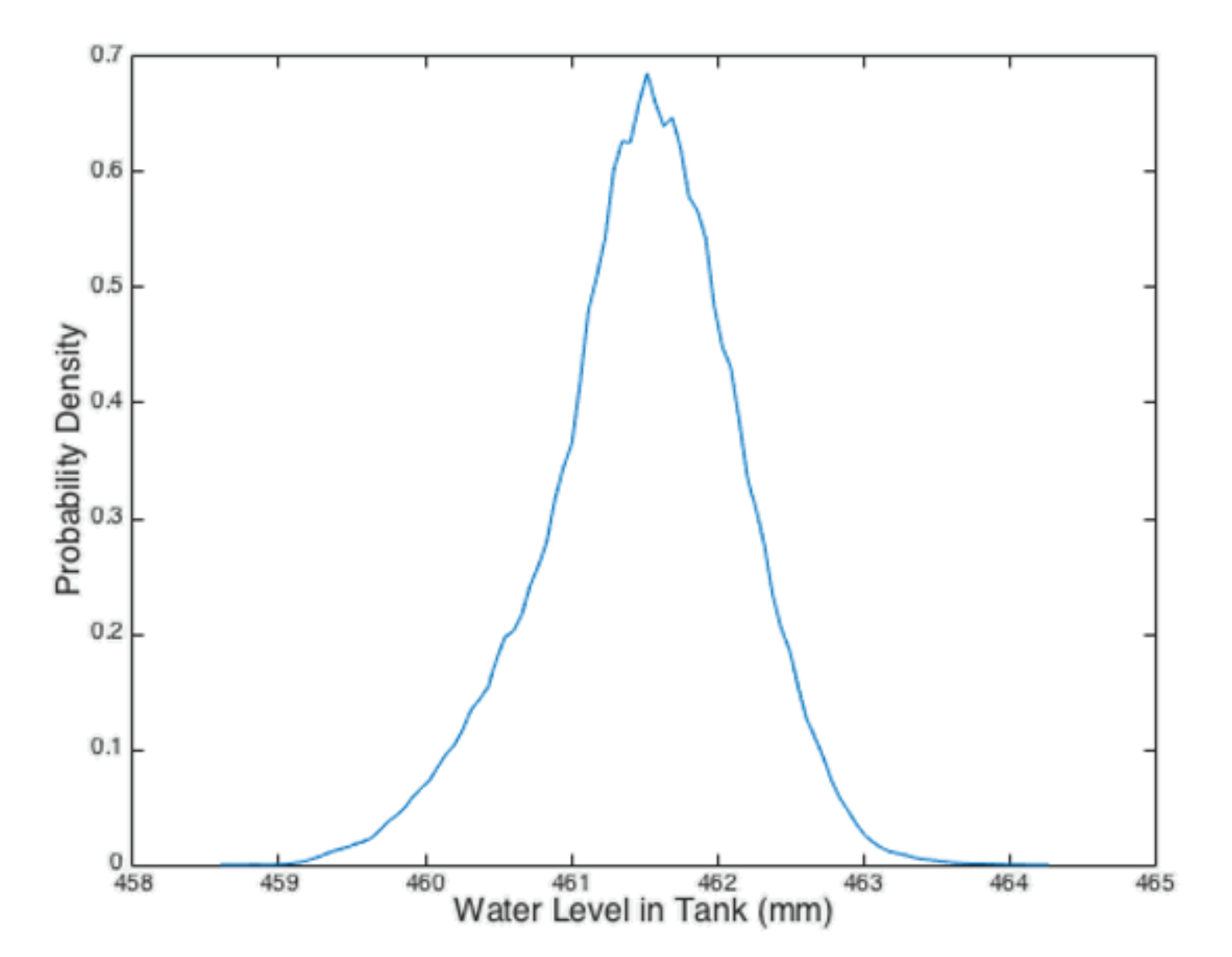}
         \caption{Noise distribution for the time series of the ultrasonic level sensor.}
         \label{fig:pdf_time_series}
     \end{figure}


\section{Visualizing the Performance}


\subsection{Noise Signal Time Series}
\acronym, as mentioned, is a sensor fingerprint based on the measurement noise from a sensor. To visualize let's consider a water level sensor in the water treatment plant. Figure~\ref{fig:time_series} shows a time series signal measured by the level sensor that is supposed to measure a constant water level in a tank. In Figure~\ref{fig:time_series} the value returned by the level sensor around a mean value is considered a noise vector. On the right-hand side, in Figure~\ref{fig:pdf_time_series} the distribution of the noise vector is shown. It is observed that the noise profile follows a Gaussian distribution. For each sensor, a fingerprint is obtained based on this noise distribution.  

\subsection{Confusion Matrix}
\begin{figure}[!htb]
\centering
\includegraphics[scale=.14]{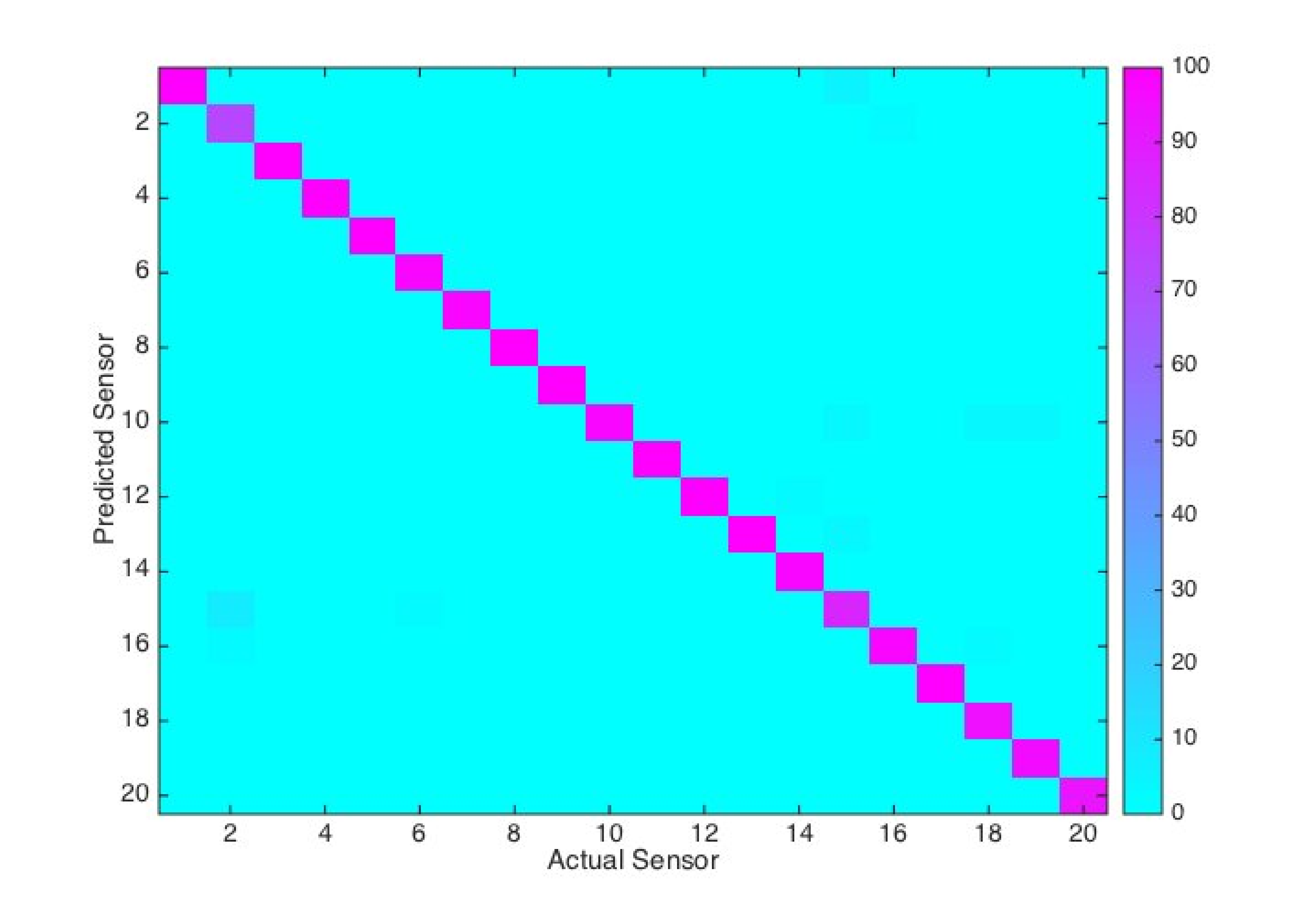}
\caption{Confusion matrix for 20 small ultrasonic sensors.}
\label{confusion_matrix}
\end{figure}

For visualizing the performance of propose \acronym, an experiment is a setup using 20 sensors of the same type and model manufactured by the same vendor. All the sensors are mounted on top of the same water tank one after another. Multiclass classification is performed by comparing each sensor with the rest of the sensors to figure out how effective is the fingerprints. In Figure~\ref{confusion_matrix} it is observed that all the sensors could be identified rightfully based on the \acronym. This result points out that for a reasonable number of sensors that is the case of a medium-scale plant, we could fingerprint sensors based on their fingerprint even for the same type of sensors. \acronym does not need any extra hardware deployment and it is a passive method for figuring out if the data is not being generated from our legitimate sensors but some malicious device or being spoofed during communicating to other devices such as PLC.  


\subsection{Limitations}
We would like to highlight that although our proposal is a passive method and it does not depend on the number of sensors under attack but there are still some limitations. 

\Paragraph{Sensor Attacks only} The proposed \acronym detects attacks on the sensors and it is not able to detect attacks on actuators. However, actuator fingerprinting techniques~\cite{raheem2016} already exist and could be used in parallel to \acronym to provide a holistic technique for attack detection.

\Paragraph{Detection Time} Stateless detection techniques (e.g., bad-data detector) or stateful detection techniques (e.g., CUSUM) might be able to raise an alarm if the attack is abrupt like a fault, but \acronym needs a chunk of data to extract the noise vector and make a decision. Therefore, \acronym might take more time in some situations as compared to other statistical detectors. However, \acronym has proven to be more successful in cases where statistical detectors had failed against a smart stealthy attacker.


\section{Summary and Conclusions}
\Paragraph{Challenges and Opportunities} 
We observed that one of the dominating challenges in CPS as compared to pure IT systems is that there is a whole lot of physical processes to be secured besides the cyber infrastructure. The same challenge of securing the physical systems becomes an opportunity if the physics of the normal process could be modeled accurately. Also, we highlighted that the integrity of data is more critical than the confidentiality of data in CPS.

\Paragraph{State of the Art} 
Attack detection is an important step toward attack mitigation and recovery. There have been extensive efforts in model-based attack detection in CPS. However, model-based attack detection techniques suffer from several limitations such as inability against stealthy and multi-point attacks, interference to the normal process.

\Paragraph{Device Fingerprinting} We put forth the idea of device fingerprinting using the hardware characteristics of sensors, such as measurement noise from a sensor. An idea called \acronym boosts the usability for being a passive (non-intrusive) attack detection solution, which is an important requirement for CPS.

\Paragraph{Conclusions} Physics-based solutions are effective in the detection of attacks to CPS. However, this approach also has its limitations. There does not exist a silver bullet to tackle all kinds of threats perfectly. Different security solutions may need to be combined to provide holistic protection for CPS.

\bibliographystyle{IEEEtran}
\bibliography{main.bib}

\begin{thebibliography}{10}
\providecommand{\url}[1]{#1}
\csname url@samestyle\endcsname
\providecommand{\newblock}{\relax}
\providecommand{\bibinfo}[2]{#2}
\providecommand{\BIBentrySTDinterwordspacing}{\spaceskip=0pt\relax}
\providecommand{\BIBentryALTinterwordstretchfactor}{4}
\providecommand{\BIBentryALTinterwordspacing}{\spaceskip=\fontdimen2\font plus
\BIBentryALTinterwordstretchfactor\fontdimen3\font minus
  \fontdimen4\font\relax}
\providecommand{\BIBforeignlanguage}[2]{{%
\expandafter\ifx\csname l@#1\endcsname\relax
\typeout{** WARNING: IEEEtran.bst: No hyphenation pattern has been}%
\typeout{** loaded for the language `#1'. Using the pattern for}%
\typeout{** the default language instead.}%
\else
\language=\csname l@#1\endcsname
\fi
#2}}
\providecommand{\BIBdecl}{\relax}
\BIBdecl

\bibitem{rajkumar_insup_CPS_intro_2010}
R.~{Rajkumar}, I.~{Lee}, L.~{Sha}, and J.~{Stankovic}, ``Cyber-physical
  systems: The next computing revolution,'' in \emph{Design Automation
  Conference}, 6 2010, pp. 731--736.

\bibitem{CPS_thesis_ETH_2018_intro}
F.~Sutton, ``\BIBforeignlanguage{en}{An efficient platform and communication
  architecture for event-triggered cyber-physical systems},'' Ph.D.
  dissertation, ETH Zurich, 2018.

\bibitem{ukranian_case2016analysis}
D.~U. Case, ``Analysis of the cyber attack on the ukrainian power grid,''
  \emph{Report}, 2016.

\bibitem{slay_miller_2008}
J.~Slay and M.~Miller, ``Lessons learned from the maroochy water breach,''
  \emph{Springer 620 US, Boston, MA}, pp. 73--82, 2008.

\bibitem{stuxnet}
N.~Falliere, L.~Murchu, and E.~Chien, ``W32 stuxnet dossier. symantec, version
  1.4,'' https://www.symantec.com/content/en/us/enterprise/media/security\_ \\
  response/whitepapers/w32\_stuxnet\_dossier.pdf, 2 2011.

\bibitem{cardenas2009challenges}
A.~Cardenas, S.~Amin, B.~Sinopoli, A.~Giani, A.~Perrig, and S.~Sastry,
  ``Challenges for securing cyber physical systems,'' in \emph{Workshop on
  future directions in cyber-physical systems security}, 2009, p.~5.

\bibitem{aurora_attack}
CNN, ``Staged cyber attack reveals vulnerability in power grid,''
  http://edition.cnn.com/ \\ 2007/US/09/26/power.at.risk/index.html, 2007.

\bibitem{German_steelmill_attack}
Wired, ``A cyberattack has caused confirmed physical damage for the second time
  ever,'' {https://www.wired.com/2015/01/german-steel-mill-hack-destruction/},
  2015.

\bibitem{Gollmann2016}
\BIBentryALTinterwordspacing
D.~Gollmann and M.~Krotofil, \emph{Cyber-Physical Systems Security}.\hskip 1em
  plus 0.5em minus 0.4em\relax Berlin, Heidelberg: Springer Berlin Heidelberg,
  2016, pp. 195--204. [Online]. Available:
  \url{https://doi.org/10.1007/978-3-662-49301-4_14}
\BIBentrySTDinterwordspacing

\bibitem{urbina_CCS2016limiting}
D.~I. Urbina, J.~A. Giraldo, A.~A. Cardenas, N.~O. Tippenhauer, J.~Valente,
  M.~Faisal, J.~Ruths, R.~Candell, and H.~Sandberg, ``Limiting the impact of
  stealthy attacks on industrial control systems,'' in \emph{Proceedings of the
  2016 ACM SIGSAC Conference on Computer and Communications Security}.\hskip
  1em plus 0.5em minus 0.4em\relax ACM, 2016, pp. 1092--1105.

\bibitem{shoukry2015}
\BIBentryALTinterwordspacing
Y.~Shoukry, P.~Martin, Y.~Yona, S.~Diggavi, and M.~Srivastava, ``Pycra:
  Physical challenge-response authentication for active sensors under spoofing
  attacks,'' in \emph{Proceedings of the 22Nd ACM SIGSAC Conference on Computer
  and Communications Security}, ser. CCS '15.\hskip 1em plus 0.5em minus
  0.4em\relax New York, NY, USA: ACM, 2015, pp. 1004--1015. [Online].
  Available: \url{http://doi.acm.org/10.1145/2810103.2813679}
\BIBentrySTDinterwordspacing

\bibitem{drone_Son2015}
\BIBentryALTinterwordspacing
Y.~Son, H.~Shin, D.~Kim, Y.~Park, J.~Noh, K.~Choi, J.~Choi, and Y.~Kim,
  ``Rocking drones with intentional sound noise on gyroscopic sensors,'' in
  \emph{Proceedings of the 24th USENIX Conference on Security Symposium}, ser.
  SEC'15.\hskip 1em plus 0.5em minus 0.4em\relax Berkeley, CA, USA: USENIX
  Association, 2015, pp. 881--896. [Online]. Available:
  \url{http://dl.acm.org/citation.cfm?id=2831143.2831199}
\BIBentrySTDinterwordspacing

\bibitem{ahmed2019state_book}
C.~M. Ahmed, J.~Zhou, and A.~P. Mathur, ``State estimation-based attack
  detection in cyber-physical systems: Limitations and solutions,''
  \emph{Cybersecurity and Privacy in Cyber Physical Systems}, p.~71, 2019.

\bibitem{yasser-2013}
S.~Yasser, M.~Paul, T.~Paulo, and S.~Mani, ``Non-invasive spoofing attacks for
  anti-lock braking systems,'' in \emph{CHES, Springer Link}, vol. 8086, 10
  2013, pp. 55--72.

\bibitem{sensor_saturationAttack_infusionpump_usenix2016}
\BIBentryALTinterwordspacing
Y.~Park, Y.~Son, H.~Shin, D.~Kim, and Y.~Kim, ``This ain{\textquoteright}t your
  dose: Sensor spoofing attack on medical infusion pump,'' in \emph{10th
  {USENIX} Workshop on Offensive Technologies ({WOOT} 16)}.\hskip 1em plus
  0.5em minus 0.4em\relax Austin, TX: {USENIX} Association, 2016. [Online].
  Available:
  \url{https://www.usenix.org/conference/woot16/workshop-program/presentation/park}
\BIBentrySTDinterwordspacing

\bibitem{sampling_race2016}
\BIBentryALTinterwordspacing
H.~Shin, Y.~Son, Y.~Park, Y.~Kwon, and Y.~Kim, ``Sampling race: Bypassing
  timing-based analog active sensor spoofing detection on analog-digital
  systems,'' in \emph{Proceedings of the 10th USENIX Conference on Offensive
  Technologies}, ser. WOOT'16.\hskip 1em plus 0.5em minus 0.4em\relax Berkeley,
  CA, USA: USENIX Association, 2016, pp. 200--210. [Online]. Available:
  \url{http://dl.acm.org/citation.cfm?id=3027019.3027037}
\BIBentrySTDinterwordspacing

\bibitem{walnut_acoustic_attack_mems_accelerometer}
T.~Trippel, O.~Weisse, W.~Xu, P.~Honeyman, and K.~Fu, ``Walnut: Waging doubt on
  the integrity of mems accelerometers with acoustic injection attacks,'' in
  \emph{2017 IEEE European Symposium on Security and Privacy (EuroS P)}, 4
  2017, pp. 3--18.

\bibitem{ahmed_QRS2017}
C.~M. Ahmed and A.~P. Mathur, ``Hardware identification via sensor
  fingerprinting in a cyber physical system,'' in \emph{2017 IEEE International
  Conference on Software Quality, Reliability and Security Companion (QRS-C)},
  7 2017, pp. 517--524.

\bibitem{Williams_purdue_reference_architecture}
\BIBentryALTinterwordspacing
T.~J. Williams, ``The purdue enterprise reference architecture,'' in
  \emph{Proceedings of the JSPE/IFIP TC5/WG5.3 Workshop on the Design of
  Information Infrastructure Systems for Manufacturing}, ser. DIISM '93.\hskip
  1em plus 0.5em minus 0.4em\relax Amsterdam, The Netherlands, The Netherlands:
  North-Holland Publishing Co., 1993, pp. 43--64. [Online]. Available:
  \url{http://dl.acm.org/citation.cfm?id=647134.716786}
\BIBentrySTDinterwordspacing

\bibitem{communication_protocol_ICS_survey_2013}
P.~Gaj, J.~Jasperneite, and M.~Felser, ``Computer communication within
  industrial distributed environment—a survey,'' \emph{IEEE Transactions on
  Industrial Informatics}, vol.~9, no.~1, pp. 182--189, 2 2013.

\bibitem{cip_protocol_odva}
ODVA, ``Common industrial protocol (cip) and the family of cip networks,''
  https://www.odva.org/Portals/0/Library/Publications\_Numbered/ \\
  PUB00123R1\_Common-Industrial\_Protocol\_and\_Family\_of\_CIP \\
  \_Networks.pdf, 2 2016.

\bibitem{enip_protocol_rockwell}
R.~Automation, ``Ethernet/ip: Industrial protocol white paper,''
  https://literature.rockwellautomation.com/idc/groups/literature/documents \\
  /wp/enet-wp001\_-en-p.pdf, 2001.

\bibitem{John_ACNS2017}
J.~H. Castellanos, D.~Antonioli, N.~O. Tippenhauer, and M.~Ochoa,
  ``Legacy-compliant data authentication for industrial control system
  traffic,'' in \emph{Applied Cryptography and Network Security}, D.~Gollmann,
  A.~Miyaji, and H.~Kikuchi, Eds.\hskip 1em plus 0.5em minus 0.4em\relax Cham:
  Springer International Publishing, 2017, pp. 665--685.

\bibitem{stuxnet_langner_2011_SnP}
R.~Langner, ``Stuxnet: Dissecting a cyberwarfare weapon,'' \emph{IEEE Security
  Privacy}, vol.~9, no.~3, pp. 49--51, 5 2011.

\bibitem{ukranian_case2017analysis_crashoverride}
\BIBentryALTinterwordspacing
US-CERT, ``Crashoverride malware,'' \emph{US-CERT Report}, 2017. [Online].
  Available: \url{https://www.us-cert.gov/ncas/alerts/TA17-163A}
\BIBentrySTDinterwordspacing

\bibitem{Saudi_aramco_marina_fireeye_2017_triton}
B.~Johnson, D.~Caban, M.~Krotofil, D.~Scali, N.~Brubaker, and C.~Glyer,
  ``Attackers deploy new ics attack framework “triton” and cause
  operational disruption to critical infrastructure,'' \emph{Threat Research
  Blog}, 2017.

\bibitem{norsk_hydro_2019_attack_ransomware}
L.~Henrik, S.~Peter, R.~Dennis, B.~Andres, and H.~Kristine, ``Attack against
  norsk hydro,'' \emph{Media Report}, 2019.

\bibitem{asco_aircraft_2019_attack_ransomware}
Z.~Zorz, ``Ransomware disrupts worldwide production for belgian aircraft parts
  maker,'' https://www.helpnetsecurity.com/2019/06/13/asco-ransomware-attack/,
  2019.

\bibitem{ghosttalk_2013}
D.~F. Kune, J.~Backes, S.~S. Clark, D.~Kramer, M.~Reynolds, K.~Fu, Y.~Kim, and
  W.~Xu, ``Ghost talk: Mitigating emi signal injection attacks against analog
  sensors,'' in \emph{2013 IEEE Symposium on Security and Privacy}, 5 2013, pp.
  145--159.

\bibitem{sonic_gun_blackhat2017}
\BIBentryALTinterwordspacing
Z.~Wang, K.~Wang, B.~Yang, S.~Li, and A.~Pan, ``Sonic gun to smart devices,''
  \emph{Blackhat USA}, 2017. [Online]. Available:
  \url{https://www.blackhat.com/docs/us-17/thursday/us-17-Wang-Sonic-Gun-To-Smart-Devices-Your-Devices-Lose
  \\ -Control-Under-Ultrasound-Or-Sound.pdf}
\BIBentrySTDinterwordspacing

\bibitem{trickorheat_kevinfu_temp_sensor_attacks_EMI_2019}
\BIBentryALTinterwordspacing
Y.~Tu, S.~Rampazzi, B.~Hao, A.~Rodriguez, K.~Fu, and X.~Hei, ``Trick or heat?
  attack on amplification circuits to abuse critical temperature control
  systems,'' \emph{CoRR}, vol. abs/1904.07110, 2019. [Online]. Available:
  \url{http://arxiv.org/abs/1904.07110}
\BIBentrySTDinterwordspacing

\bibitem{marco_cpdy2016}
\BIBentryALTinterwordspacing
M.~Rocchetto and N.~O. Tippenhauer, ``{CPDY:} extending the dolev-yao attacker
  with physical-layer interactions,'' \emph{CoRR}, vol. abs/1607.02562, 2016.
  [Online]. Available: \url{http://arxiv.org/abs/1607.02562}
\BIBentrySTDinterwordspacing

\bibitem{eireann_2013_greyhat_PLC_vulnerabiity}
\BIBentryALTinterwordspacing
E.~Leverett and R.~Wightman, ``Vulnerability inheritance in programmable logic
  controllers,'' \emph{US-CERT Report}, 2013. [Online]. Available:
  \url{https://ics-cert.us-cert.gov/content/cyber-threat-source-descriptions}
\BIBentrySTDinterwordspacing

\bibitem{ruben_backdoors_PLC_2012_blackhat}
\BIBentryALTinterwordspacing
R.~Santamarta, ``Here be backdoors: A journey into the secrets of industrial
  firmware.'' \emph{CoRR}, 2012. [Online]. Available:
  \url{https://media.blackhat.com/bh- us- 12/Briefings/ Santamarta/BH US 12
  Santamarta \\ Backdoors WP.pdf}
\BIBentrySTDinterwordspacing

\bibitem{fovino_2009_malware_PLC}
\BIBentryALTinterwordspacing
I.~N. Fovino, A.~Carcano, M.~Masera, and A.~Trombetta, ``An experimental
  investigation of malware attacks on {SCADA} systems,'' \emph{International
  Journal of Critical Infrastructure Protection}, vol.~2, no.~4, pp. 139 --
  145, 2009. [Online]. Available:
  \url{http://www.sciencedirect.com/science/article/pii/S1874548209000419}
\BIBentrySTDinterwordspacing

\bibitem{robert_turk_2005_PLC_DoS}
\BIBentryALTinterwordspacing
R.~J. Turk, ``Cyber incidents involving control systems,'' 2005. [Online].
  Available:
  \url{https://pdfs.semanticscholar.org/1f8f/a134eca5fe92143bd154ec9f6446b38b63ae.pdf}
\BIBentrySTDinterwordspacing

\bibitem{Anand_ESORICS2017_PLC_ladderlogicbomb}
\BIBentryALTinterwordspacing
N.~Govil, A.~Agrawal, and N.~O. Tippenhauer, ``On ladder logic bombs in
  industrial control systems,'' \emph{CoRR}, vol. abs/1702.05241, 2017.
  [Online]. Available: \url{http://arxiv.org/abs/1702.05241}
\BIBentrySTDinterwordspacing

\bibitem{MS_thesis_PLC_attack_NTNU_2013}
X.~Morten~Gjendemsj{\o}, ``Creating a weapon of mass disruption: Attacking
  programmable logic controllers,'' Ph.D. dissertation, Norwegian University of
  Science and Technology, 6 2013.

\bibitem{swat2016}
A.~P. Mathur and N.~O. Tippenhauer, ``Swat: a water treatment testbed for
  research and training on ics security,'' in \emph{2016 International Workshop
  on Cyber-physical Systems for Smart Water Networks (CySWater)}, 4 2016, pp.
  31--36.

\bibitem{Rizwan_ESORICS2017_stealthyAtt}
Q.~R., M.~C.~A. C.M., and R.~J., ``Multistage downstream attack detection in a
  cyber physical system,'' in \emph{CyberICPS Workshop 2017, in conjunction
  with ESORICS 2017}, 9 2017.

\bibitem{Carlos_Justin_CDC2016_stealthyAtt}
C.~Murguia and J.~Ruths, ``Characterization of a cusum model-based sensor
  attack detector,'' in \emph{2016 IEEE 55th Conference on Decision and Control
  (CDC)}, 12 2016, pp. 1303--1309.

\bibitem{pasqualetti2013attack_control4}
F.~Pasqualetti, F.~D{\"o}rfler, and F.~Bullo, ``Attack detection and
  identification in cyber-physical systems,'' \emph{IEEE transactions on
  automatic control}, vol.~58, no.~11, pp. 2715--2729, 2013.

\bibitem{carlos_msc2016}
C.~Murguia and J.~Ruths, ``Cusum and chi-squared attack detection of
  compromised sensors,'' in \emph{2016 IEEE Conference on Control Applications
  (CCA)}, 9 2016, pp. 474--480.

\bibitem{liu_ccs2009_first_work_}
\BIBentryALTinterwordspacing
Y.~Liu, P.~Ning, and M.~K. Reiter, ``False data injection attacks against state
  estimation in electric power grids,'' in \emph{Proceedings of the 16th ACM
  Conference on Computer and Communications Security}, ser. CCS ’09.\hskip
  1em plus 0.5em minus 0.4em\relax New York, NY, USA: Association for Computing
  Machinery, 2009, p. 21–32. [Online]. Available:
  \url{https://doi.org/10.1145/1653662.1653666}
\BIBentrySTDinterwordspacing

\bibitem{Ahmed_AsiaCCS2017_stealthyAtt}
\BIBentryALTinterwordspacing
C.~M. Ahmed, C.~Murguia, and J.~Ruths, ``Model-based attack detection scheme
  for smart water distribution networks,'' in \emph{Proceedings of the 2017 ACM
  on Asia Conference on Computer and Communications Security}, ser. ASIA CCS
  '17.\hskip 1em plus 0.5em minus 0.4em\relax New York, NY, USA: ACM, 2017, pp.
  101--113. [Online]. Available:
  \url{http://doi.acm.org/10.1145/3052973.3053011}
\BIBentrySTDinterwordspacing

\bibitem{wang2014srid_ESORICS2014_invariants}
Y.~Wang, Z.~Xu, J.~Zhang, L.~Xu, H.~Wang, and G.~Gu, ``Srid: State relation
  based intrusion detection for false data injection attacks in scada,'' in
  \emph{European Symposium on Research in Computer Security}.\hskip 1em plus
  0.5em minus 0.4em\relax Springer, 2014, pp. 401--418.

\bibitem{adepuMathurIFIPSEC2016_invariants}
S.~Adepu and A.~Mathur, ``Using process invariants to detect cyber attacks on a
  water treatment system,'' in \emph{Proceedings of the 31st International
  Conference on ICT Systems Security and Privacy Protection - IFIP SEC 2016
  (IFIP AICT series)}.\hskip 1em plus 0.5em minus 0.4em\relax Springer, 2016.

\bibitem{choi2018detecting_dongyan_ccs2018_drone_control_invariant}
H.~Choi, W.-C. Lee, Y.~Aafer, F.~Fei, Z.~Tu, X.~Zhang, D.~Xu, and X.~Deng,
  ``Detecting attacks against robotic vehicles: A control invariant approach,''
  in \emph{Proceedings of the 2018 ACM SIGSAC Conference on Computer and
  Communications Security}, 2018, pp. 801--816.

\bibitem{bruno2015_watermarking}
Y.~Mo, S.~Weerakkody, and B.~Sinopoli, ``Physical authentication of control
  systems: Designing watermarked control inputs to detect counterfeit sensor
  outputs,'' \emph{IEEE Control Systems Magazine}, vol.~35, no.~1, pp. 93--109,
  Feb 2015.

\bibitem{bai2014kalman_control1}
C.-Z. Bai and V.~Gupta, ``On kalman filtering in the presence of a compromised
  sensor: Fundamental performance bounds,'' in \emph{2014 American control
  conference}.\hskip 1em plus 0.5em minus 0.4em\relax IEEE, 2014, pp.
  3029--3034.

\bibitem{CPSweek2016_stealthy_replayATT}
C.~M. Ahmed, S.~Adepu, and A.~Mathur, ``Limitations of state estimation based
  cyber attack detection schemes in industrial control systems,'' in \emph{2016
  Smart City Security and Privacy Workshop (SCSP-W)}, 4 2016, pp. 1--5.

\bibitem{shoukry2018smt_control3}
Y.~Shoukry, M.~Chong, M.~Wakaiki, P.~Nuzzo, A.~Sangiovanni-Vincentelli, S.~A.
  Seshia, J.~P. Hespanha, and P.~Tabuada, ``Smt-based observer design for
  cyber-physical systems under sensor attacks,'' \emph{ACM Transactions on
  Cyber-Physical Systems}, vol.~2, no.~1, pp. 1--27, 2018.

\bibitem{mishra2016secure_estimation_control2}
S.~Mishra, Y.~Shoukry, N.~Karamchandani, S.~N. Diggavi, and P.~Tabuada,
  ``Secure state estimation against sensor attacks in the presence of noise,''
  \emph{IEEE Transactions on Control of Network Systems}, vol.~4, no.~1, pp.
  49--59, 2016.

\bibitem{noisematters_Mujeeb_acsac2018}
\BIBentryALTinterwordspacing
C.~M. Ahmed, J.~Zhou, and A.~P. Mathur, ``Noise matters: Using sensor and
  process noise fingerprint to detect stealthy cyber attacks and authenticate
  sensors in cps,'' in \emph{Proceedings of the 34th Annual Computer Security
  Applications Conference}, ser. ACSAC '18.\hskip 1em plus 0.5em minus
  0.4em\relax New York, NY, USA: ACM, 2018, pp. 566--581. [Online]. Available:
  \url{http://doi.acm.org/10.1145/3274694.3274748}
\BIBentrySTDinterwordspacing

\bibitem{krotofilLarsenGollman_process_matters}
M.~Krotofil, J.~Larsen, and D.~Gollmann, ``The process matters: Ensuring data
  veracity in cyber-physical systems,'' in \emph{Proceedings of the 10th ACM
  Symposium on Information, Computer and Communications Security}, 2015, pp.
  133--144.

\bibitem{roth2016physical_authentication_BruceMcmillan}
T.~Roth and B.~McMillin, ``Physical attestation in the smart grid for
  distributed state verification,'' \emph{IEEE Transactions on Dependable and
  Secure Computing}, vol.~15, no.~2, pp. 275--288, 2016.

\bibitem{sunjun_ieee_snp_2018_dataset_swat}
Y.~Chen, C.~M. Poskitt, and J.~Sun, ``Learning from mutants: Using code
  mutation to learn and monitor invariants of a cyber-physical system,'' in
  \emph{2018 IEEE Symposium on Security and Privacy (SP)}.\hskip 1em plus 0.5em
  minus 0.4em\relax IEEE, 2018, pp. 648--660.

\bibitem{Anand_agrawal2018poster_AsiaCCS2018}
A.~Agrawal, C.~M. Ahmed, and E.-C. Chang, ``Poster: Physics-based attack
  detection for an insider threat model in a cyber-physical system,'' in
  \emph{Proceedings of the 2018 on Asia Conference on Computer and
  Communications Security}, 2018, pp. 821--823.

\bibitem{kohno2005}
T.~Kohno, A.~Broido, and K.~C. Claffy, ``Remote physical device
  fingerprinting,'' \emph{IEEE Transactions on Dependable and Secure
  Computing}, vol.~2, no.~2, pp. 93--108, 4 2005.

\bibitem{moon1999}
S.~B. Moon, P.~Skelly, and D.~Towsley, ``Estimation and removal of clock skew
  from network delay measurements,'' in \emph{INFOCOM '99. Eighteenth Annual
  Joint Conference of the IEEE Computer and Communications Societies.
  Proceedings. IEEE}, vol.~1, 3 1999, pp. 227--234 vol.1.

\bibitem{paxson1998}
\BIBentryALTinterwordspacing
V.~Paxson, ``On calibrating measurements of packet transit times,'' in
  \emph{Proceedings of the 1998 ACM SIGMETRICS Joint International Conference
  on Measurement and Modeling of Computer Systems}, ser. SIGMETRICS
  '98/PERFORMANCE '98.\hskip 1em plus 0.5em minus 0.4em\relax New York, NY,
  USA: ACM, 1998, pp. 11--21. [Online]. Available:
  \url{http://doi.acm.org/10.1145/277851.277865}
\BIBentrySTDinterwordspacing

\bibitem{raheem2015}
S.~V. Radhakrishnan, A.~S. Uluagac, and R.~Beyah, ``Gtid: A technique for
  physical device and device type fingerprinting,'' \emph{IEEE Transactions on
  Dependable and Secure Computing}, vol.~12, no.~5, pp. 519--532, 9 2015.

\bibitem{dey-2014}
S.~Dey, N.~Roy, W.~Xu, R.~R. Choudhury, and S.~Nelakuditi, ``Accelprint:
  Imperfections of accelerometers make smartphones trackable,'' in
  \emph{Network and Distributed System Security Symposium (NDSS)}, 2014.

\bibitem{raheem2016}
D.~Formby, P.~Srinivasan, A.~Leonard, J.~Rogers, and R.~Beyah, ``Who's in
  control of your control system? device fingerprinting for cyber-physical
  systems,'' in \emph{NDSS}, 4 2016.

\bibitem{NoiSense_eprint_Mujeeb}
C.~{Mujeeb Ahmed}, A.~{Mathur}, and M.~{Ochoa}, ``{NoiSense: Detecting Data
  Integrity Attacks on Sensor Measurements using Hardware based
  Fingerprints},'' \emph{ArXiv e-prints}, 12 2017.

\bibitem{gerdes2006}
R.~M. Gerdes, T.~E. Daniels, M.~Mina, and S.~F. Russell, ``Device
  identification via analog signal fingerprinting: A matched filter approach,''
  in \emph{NDSS}, 2006.

\bibitem{wadi2017}
\BIBentryALTinterwordspacing
C.~M. Ahmed, V.~R. Palleti, and A.~P. Mathur, ``Wadi: A water distribution
  testbed for research in the design of secure cyber physical systems,'' in
  \emph{Proceedings of the 3rd International Workshop on Cyber-Physical Systems
  for Smart Water Networks}, ser. CySWATER '17.\hskip 1em plus 0.5em minus
  0.4em\relax New York, NY, USA: ACM, 2017, pp. 25--28. [Online]. Available:
  \url{http://doi.acm.org/10.1145/3055366.3055375}
\BIBentrySTDinterwordspacing

\bibitem{jenny-2013}
J.~T., E.~T., R.~N., and M.~A, \emph{Ultrasonic Fluid Quantity Measurement in
  Dynamic Vehicular Applications: A Support Vector Machine Approach}.\hskip 1em
  plus 0.5em minus 0.4em\relax Springer, 2013.

\bibitem{coutard-2005}
F.~Coutard, E.~Tisserand, and P.~Schweitzer, ``The temperature influence on the
  piezoelectric transducer noise, measurements and modelling,'' in \emph{IEEE
  Ultrasonics Symposium}, vol.~3, 2005, pp. 1652--1655.

\bibitem{petr-2011}
S.~Petr, M.~Jiri, and S.~Josef, ``Noise in piezoelectric ceramics at the low
  temperature,'' in \emph{Radio Engineering}, vol.~20, 2011.

\bibitem{flotech_radar}
FloTech, ``{RD700} 2-wire radar level transmitter,''
  {http://www.flotech.com.sg/ \\ downloads/rd700-radar-level-transmitter.pdf},
  2016.

\bibitem{flotech_flowmeter}
Flotech, ``Electromagnetic flowmeter,'' http://www.unhas.ac.id/rhiza/arsip/ \\
  iwormee2009/old-archieve/Spec\%20FIT.pdf, 2016.

\bibitem{accuracy_radar}
Indumart, ``Accuracy of the radar measurements,''
  http://www.indumart.com/Level-measurement-3.pdf, 2012.

\bibitem{antenna_pattern}
F.~Ustuner, E.~Aydemir, E.~Guleç, M.~Ilarslan, M.~Celebi, and E.~Demirel,
  ``Antenna radiation pattern measurement using an unmanned aerial vehicle
  (uav),'' in \emph{2014 XXXI URSI General Assembly and Scientific Symposium
  (URSI GASS)}, 8 2014, pp. 1--4.

\bibitem{flowmeter2006}
D.~Lincoln, ``An investigation into an electromagnetic flowmeter for use with
  low conductivity liquids i,'' 9 2006.

\end{thebibliography}

\end{document}